\documentclass{WileyMSP-template}
 \usepackage{comment}

\usepackage{color}
\usepackage{bm}

\begin{document}
\pagestyle{fancy}
%\rhead{\includegraphics[width=2.5cm]{vch-logo.png}}

\title{Magnetic Transparent Conductors for Spintronic Applications}

\maketitle

\author{Pino D'Amico*,} 
\author{Alessandra Catellani,}
\author{Alice Ruini,}
\author{Stefano Curtarolo,}
\author{Marco Fornari,}
\author{Marco Buongiorno Nardelli,} 
\author{Arrigo Calzolari**}

\begin{affiliations}
P. D'Amico*, A. Catellani, A. Calzolari** \\
{\it \small Istituto Nanoscienze CNR-NANO-S3, I-4115 Modena, Italy}\\
*Email: pino.damico@nano.cnr.it
**Email: arrigo.calzolari@nano.cnr.it

A. Ruini \\
{\it \small Dipartimento di Fisica, Informatica e
Matematica, Universit\'a di Modena e Reggio Emilia, I-41125
Modena, Italy}

S. Curtarolo\\
{\it \small Department of Materials Science and Engineering, Duke University, Durham, NC 27708, USA}

M. Fornari\\
{\it \small Department of Physics, Central Michigan University, Mt. Pleasant, MI 48859}

M. Buongiorno Nardelli\\
{\it \small Department of Physics, University of North Texas, Denton, TX 76203, USA}
\end{affiliations}

\keywords{Transparent conductors, magnetic materials, high-throughput simulations, DFT}

\begin{abstract}
 Transparent Conductors (TCs) exhibit optical transparency and electron conductivity, and are essential for many opto-electronic and photo-voltaic devices. The most common TCs are electron-doped oxides, which have few limitations when transition metals are used as dopants. Non-oxides TCs have the potential of extending the class of materials to the magnetic realm, bypass technological bottlenecks, and bring TCs to the field of spintronics. Here we propose new functional materials that combine transparency and conductivity with magnetic spin polarization that can be  used for spintronic applications, such as spin filters.

By using high-throughput first-principles techniques, we identified a large number of potential TCs, including non-oxides materials. Our results indicate that proper doping with transition metals introduces a finite magnetization that can provide spin filtering up to 90\% in the electrical conductivity, still maintaining a transparency greater than 90\%. 
\end{abstract}

%%%%%%%%%%%%%%%%%%%%%%%%
\section{Introduction}
%%%%%%%%%%%%%%%%%%%%%%%%

\begin{comment}
%%%%%%%
\AlC{Alessandra Catellani, ORCID 0000-0001-5197-7186}

\AR{Alice Ruini, ORCID 0000-0002-7987-1858}

\SC{Stefano Curtarolo, ORCID 0000-0003-0570-8238}

\MF{Marco Fornari, ORCID 0000-0001-6527-8511}

\MBN{Marco Buongiorno Nardelli, ORCID 0000-0003-0793-5055}

\ArC{Arrigo Calzolari, ORCID 0000-0002-0244-7717}
\end{comment}

%%%%%%%%%%%%%%%%%%%%%%%%%%%%%%%%%%%%%%%%%%%%%%%
%%%%%%%%%%%%%%%%%%%%%%%%%%%%%%%%%%%%%%%%%%%%%%%
Transparent conductors (TCs) are unique materials that simultaneously exhibit optical transparency ($\sim 90\%$) and high electrical conductivity ($\rho\sim 10^{-4} \Omega \cdot cm$). This unusual combination makes TCs essential for the realization of electromagnetic shielding layers and transparent electrodes, which constitute the building block for optoelectronic devices in solar energy applications \cite{Granqvist07}, touch-screens \cite{wu18}, flat panel displays \cite{Chae_2001}, smart electrochromic windows \cite{Runnerstrom14}, plasmonics \cite{naik2013}, and organic electronics \cite{ellmer2012}.
The diverse portfolio of applications demands materials that are able to meet the structural, electro-magnetic, and chemical variability as well as considerations regarding manufacturing, integration in complex device architectures, environmental compatibility, and cost. These needs may be met by high-throughput (HT) analyses designed to investigate large sets of candidate materials.

The most common materials used as TCs are heavily-doped wide-bandgap metal oxides (TCOs), such as indium tin oxide (Sn:In$_2$O$_3$, ITO), aluminum zinc oxide (Al:ZnO, AZO), and fluorine tin oxide (F:SnO$_2$, FTO) \cite{Minami05}. TCOs are n-type degenerate semiconductors, resulting from the transfer of electrons from the dopants to the conduction band of the oxide host. The electronic transport properties can be partially controlled  by varying the dopant concentration. Upon doping, the optical bandgap shifts towards higher energy, as described by the Burnstein-Moss model \cite{moss54,calzolari2014}, preserving the original transparency of the  metal-oxide host. ITO, for example, combines the highest electrical conductivity ($\rho = 7.2 \times 10^{-4}~\Omega \cdot$cm, $n_{el} \simeq [10^{20} - 10^{21}$] cm$^{-3}$) \cite{Chen2013,Lin14} and  about  90\%  optical transparency  in  the  visible  range \cite{Swallow19}. The realization of p-type TCOs remains instead much more challenging, due to the lower mobility of the hole carriers in metal oxides \cite{Hautier13,Woods18}. 
Similarly, the extension of transparent conductivities to magnetic materials has not been solved. If achieved, the realization of magnetic TCs would open up the integration into transparent spintronic applications \cite{Toyosaki04} and spin manipulation through electronic and/or optical interactions, such as field effect, spin filtering and photo carrier injection \cite{Panda19}. Doping  metal-oxides with magnetic transition metal (TM) elements would seem the easiest way to impart  a magnetic character to TCs. 
 However, the high chemical affinity between d-orbitals of TMs and oxygen states often results in localized chemical bonds and the emergence of mid-gap states (i.e. no free charge donation), which are detrimental for both the conductivity (carrier traps) and the transparency (optical interband transitions) of the system. This is the case, for instance,  of Fe:ZnO \cite{Srinivasulu17}, and Mn:ZnO \cite{Ahmed17}. 
 In other cases, e.g. Cr:In2O3 \cite{Farvid12}, oxygen vacancies and features of the microstructure (interfaces between crystallites and grains) are responsible for quenching the ferromagnetic order in TM-doped metal oxides. 
 Furthermore, the exchange splitting may enhance the transitions between partially occupied TM d-states giving rise to near-IR absorption.
Thus, even though a few attempts of obtaining magnetic TCOs have been proposed (e.g. Mo:In$_2$O$_3$ \cite{Medvedeva2006}), at present magnetism in TCOs remains elusive.

To overcome the limitations in TCOs, non-oxides materials should be considered.
The most promising non-oxide TCs include disordered and low-dimensional systems such as carbon-based  materials (e.g. nanotubes  and  graphene) \cite{Wu14,Zhang22,WASSEI201052}, ultra-small metallic  nanoparticles (e.g Ag)\cite{C1NR10048C},  and  conductive  polymers (e.g. PEDOT:PSS) \cite{Lee18,Dauzon20}.  
While these materials have the advantage to be cheap, easy to synthesize and mostly flexible, their conduction properties (both mobility and charge density) remain more than one order of magnitude smaller than the crystalline counterpart. In addition, their integration in current manufacturing processes is challenging.

In this paper, 
we propose a HT investigation, based on first principles simulations, for the screening and the design of novel TCs that are simultaneously crystalline, non-oxide, and magnetic.
HT screening is one of the most powerful methodologies in computational materials science \cite{aflowlibPAPER2023,aflowPAPER2023, supka2017, Calderon_cms_2015, Jain20112295, Pizzi2016218};
representative but incomplete list of examples of the impact of HT methods is given by the discovery of novel topological insulators \cite{curtarolo:TIs}, 2D materials \cite{Mounetarxiv2017}, and p-type conductors \cite{Hautier13, hautier2014}. 
We identify a set of specific physical descriptors, \cite{Curtarolo2013, Fornari2013} 
%\MF{to be added: Nature Materials 12, 191 (2013) and Physics 6, 140 (2013)}
then we search within the AFLOW repository \cite{aflowlibPAPER2023,aflowPAPER2023} for potential crystalline hosts, which may act as  TCs upon doping.
After having classified the potential materials according to symmetry, gap-nature, and other relevant properties, we select a set of promising non-oxide materials to explore the effect of systematically doping them with the entire set of 30 TMs.
This workflow leads to a set of crystalline TCs with a finite magnetization that exhibit a net spin-polarized electron conductivity and can be exploited for spin filters applications. 
These materials open a new route to magnetic optoelectronic devices, such as transparent spintronics or all-optically controlled spin-systems.

%%%%%%%%%%%%%%%%%%%%%%%%%%%%%%%%%%%%%%%%%%%%%%%%%%%%%%%%%%%%%%%%%%%%%%%%%%
\section{Descriptors and screening} \label{ht-searches}
%%%%%%%%%%%%%%%%%%%%%%%%%%%%%%%%%%%%%%%%%%%%%%%%%%%%%%%%%%%%%%%%%%%%%%%%%%

The first step in our discovery procedure is the identification of the physical descriptors that characterize good TC materials. 
The descriptors are used to select promising TCs included in the {AFLOW} repository \cite{aflowlibPAPER2023,aflowPAPER2023}. The whole dataset contains over 3.5 millions compounds characterized in terms of many electronic structure properties computed at the level of density functional theory (DFT).

%%%%%%%%%%%%%%%%%%%%%%%%%%%%%%%%%%%%%%%%%%%%%%%
%%%%%%%%%%%%%%%%%%%%%%%%%%%%%%%%%%%%%%%%%%%%%%%
\begin{figure}[htb]
\begin{center}
    \includegraphics[width=0.7\columnwidth]{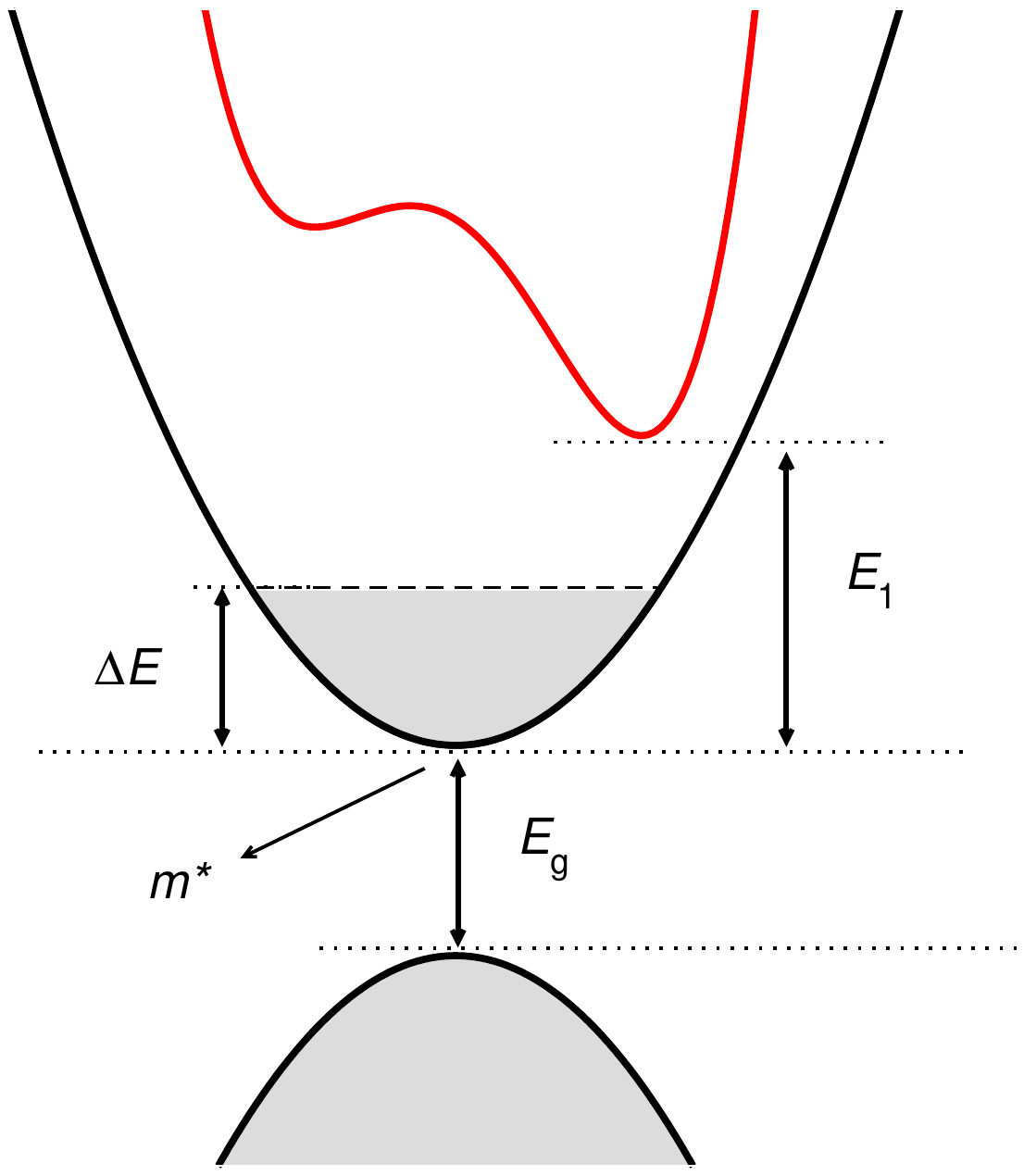}
    \vspace{5mm}
    \caption{\small Sketch of a prototype band structure highlighting the essential parameters used as descriptors to distill promising materials from the {AFLOW} database. $E_g$ is the energy gap of the pristine material; $E_1$ is the energy distance between the bottom of the first and the second conduction bands;  $\Delta$E is the Fermi energy shift of the doped material measured with respect to the bottom of the first conduction band;  m$^*$ is the effective electron mass. Gray areas identify the occupied bands of the doped system.
    }
    \label{SKETCH}
  \end{center}
\end{figure}

%%%%%%%%%%%%%%%%%%%%%%%%%%%%%%%%%%%%%%%%%%%%%%%
%%%%%%%%%%%%%%%%%%%%%%%%%%%%%%%%%%%%%%%%%%%%%%%
    
In the definition of the descriptors, we considered ITO and AZO as guiding prototypes to achieve transparency and conductivity. 
In both cases, the main effect of the substitutional doping is the occupation of electronic states at the the bottom of the conduction band \cite{calzolari2014}. 
The structural features of the host lattice are only slightly perturbed and no dangling bonds or states in the pristine host gap are formed. 
We focus on the electronic band structure of the host material in order to determine the values of the descriptors.
Figure \ref{SKETCH} summarizes the band-structure features that characterize
the properties of the doped semiconductor from the point of view of transparent conductivity.
First, the host must be transparent, at least in the visible range; this implies an energy bandgap $E_g \ge 3.1$ eV.
%In view of 
Because of the underestimation of the bandgap by standard DFT, the bandgap reported in the {AFLOW} database could be much smaller (up to 30\%) than the experimental ones. For this reason, we fixed the first query criterion as
\begin{equation}
E_g > 1.0~{\textrm eV}.
\label{eq1}
\end{equation}
The second descriptor is the electron effective mass $m^*$ of the first conduction band: assuming that after the n-doping  a charge density is injected in the conduction band of the host material, 
the effective mass  is representative of a good or a bad conductor. 
Best conducting TCs, such as ITO and AZO, have effective masses in the range $m^*=0.2-0.5$ $m_0$,  where $m_0$ is the free electron mass \cite{GUPTA198933,Lu2007}. Thus, we set the second criterion as: 
\begin{equation}
m^* < \frac{1}{2} m_0,
\label{eq2}
\end{equation}
where $m^*$ is evaluated by a parabolic fit as the curvature of the first conduction band around the minimum. 
The effect of charge injection is a shift of the Fermi level ($E_{\textrm F}$) from midgap (undoped insulator)  to the bottom of the conduction band (degenerate semiconductor). We labelled $\Delta E$ the energy position of $E_{\textrm F}$ with respect to the 
 conduction band minimum. $\Delta E$ is an indicator of the electron charge $n_{el}$ available for transport. 

The electrons in the conduction band can undergo optical transitions to higher energy bands. In order to ensure that transparency is not lost after doping, we require the transition energy ($E_1$) between the first and the second conduction band to be large enough when compared with the energy of  light in the visible spectrum. Our third descriptor is
\begin{equation}
E_1 > 2.0~eV.
\label{eq3}
\end{equation}
The last requirement is made under the assumption that the $\Delta E$ lies in a region where the parabolic approximation is still valid.
No conditions are instead imposed on the chemical composition of the materials. 
The descriptors expressed in Eqs.~(\ref{eq1}-\ref{eq3}) were  used to scan the full {AFLOW} database.
The screening procedure identifies 115 potential TC materials, that we report in Tables S1-S4 of Supporting Information (SI). Those materials were classified according to symmetry, oxygen presence, and relevant physical properties. All systems are non-magnetic.
The list includes binary, ternary, and quaternary compounds, whose composition spans the entire periodic table. 
Interestingly, the total list correctly includes also ZnO and In$_2$O$_3$ that are the archetypes of TCOs. This confirms, {\em a posteriori}, the appropriate choice of the physical descriptors. 
Since we are interested in non-oxide materials, we discarded all entries that include oxygen as a component. This restricted 
the choice to the 39 candidates reported in Table \ref{tableI}.
Most systems are binary compounds and include TM-halides, nitrides, alkali-halides and alkali-chalcogenides. Most ternary systems are composed of alkali and halide elements, while the remaining quaternary systems have nitrogen-halide composition.
The majority of the screened crystals have cubic or tetragonal symmetry, both direct and indirect bandgap materials are in the list.  
Silver halides (AgF, AgCl, AgBr) are well-known materials used in the photographic process and in electro- and photoelectrochemistry \cite{glaus03}; copper halides (CuCl and CuBr) are studied for their photochemical and photophysical properties \cite{Peng10}. Cadmium di-halide crystals (CdF$_2$, CdCl$_2$, CdBr$_2$) show unique birefringent optical characteristics \cite{kondo81}. ZnBr$_2$ is largely used in catalytic applications and rechargeable batteries \cite{Tu00}, while HgF$_2$ is used for industrial applications such as halogen metallurgy and etching semiconductor devices \cite{Hargittai09}. GaN is one of the most used wide-band gap semiconductors for both power electronic  \cite{Flack16} and optoelectronic applications \cite{Zhao22}. 
YN has high mechanical stability and it plays a key role in the  fabrication of ternary compounds with Ga or In \cite{Mancera2003}.
Alkali-metal halides (LiCl, LiBr, NaBr, NaI, RbF) are well-known stable salts, whose photoconductivity is characterized by F-centers absorption \cite{Jenkin1978}. 
Dialkali-metal monochalcogenides (Na$_2$S, Na$_2$Se, Na$_2$Te, K$_2$S, K$_2$Se,  K$_2$Te, Rb$_2$S, Rb$_2$Se) exhibit high electrical mobility even in two-dimensional layers \cite{Hua18}. Metal nitride fluorides (Mg$_2$NF) are known as {pseudo-oxides} in view of the structural similarities with the corresponding oxide crystals \cite{Brogan12},
and hexafluorostantanates (Cs$_2$SnF$_6$) are studied for their unique
phosphorescent properties \cite{Arai11}. Alkali Cadmium tetrachlorides (Rb$_2$CdCl$_4$, Cs$_2$CdCl$_4$) are stable in their ferroelastic phase and are sensitive to light polarization \cite{Wenger01}. Quaternary compounds have been predicted theoretically, but not yet synthesized. 
%%%%%%%%%%%%%%%%%%%%%%%%%%%%%%%%%%%%%%%%%%%%%%%
%%%%%%%%%%%%%%%%%%%%%%%%%%%%%%%%%%%%%%%%%%%%%%%
%
\newpage
\begin{table*}[htb]
    \caption {Non-oxide materials resulting from screenig the AFLOW database with the descriptors in Eqs.~(\ref{eq1}-\ref{eq3}). List contains information on  number of constituents atoms,  symmetry group,  Hubbard parameter $U$, effective mass ($m^*/m_0$), gap type and its calculated ($E_g^{Th}$),  and experimental ($E_g^{Exp}$) values available in literature \cite{glaus03, Strehlow73}.}
   \begin{center}
    \begin{tabular}{c|c|c|c|c|c|c|c}
     \hline\hline
 			  & Material	    & Space Group       & U (eV) & $m^*/m_0$ & Gap   & $E_g^{Th}$ (eV)& $E_g^{Exp}$ (eV)\\
    \hline\hline
\bf{Binary} 	&	AgF 	        &  {\em FM-3m}      &	6.95, 3.27 	            &	0.41    &	I 	& 1.61  & 2.80\\			
			&	AgCl		    &  {\em FM-3m}      &	8.09, 3.02	        	&	0.26    &	I 	& 2.57  & 3.25\\			
			&	AgBr 	        &  {\em FM-3m}	    &	9.04, 3.18	         	&	0.23    &	I 	& 2.53  & 2.68\\	
			&	CuCl	      	&  {\em F-43m}      &	10.00, 1.64	            &	0.33    &	D 	& 2.92  & 3.33\\			
TM-			    &	CuBr		    &  {\em F-43m}      &	9.03, 2.01	         	&	0.24    &	D 	& 2.80  & 2.99\\
Halides		    &	ZnBr$_2$	    &  {\em R-3m} 	    &	12.71, 5.48            	& 	0.13    & 	I 	& 4.95  & 3.50\\		
			&	CdF$_2$	        &  {\em Fm-3m}      &	9.30, 9,25              & 	0.41    & 	I 	& 6.20  & 6.00\\		
			&	CdCl$_2$	    &  {\em R-3m} 	    &	9.69, 6.20           	& 	0.21    & 	I 	& 5.53  & 5.70\\		
			&	CdBr$_2$	    &  {\em R-3m} 	    &	9.78, 5,50           	& 	0.15    & 	I 	& 4.45  & 4.47\\		
			&	HgF$_2$	        &  {\em Fm-3m}	    &	5.99, 6.90            	& 	0.31    & 	D 	& 3.20  & 5.54\\
\hline
Nitrides		&	GaN		        &  {\em P6$_{3}$mc} &   19.18, 3.92	            & 	0.34    & 	D 	& 2.83  & 3.24\\
			&	YN		        &  {\em F-43m}      &	0.13, 3.17 	            & 	0.34    & 	I  	& 3.08  & 1.90\\
\hline
			&	LiCl		    &  {\em Fm-3m}      &	0.08, 6.46  	        & 	0.48    & 	D 	& 8.65  & 9.33\\
			&	LiBr		    &  {\em Fm-3m}      &	0.02, 5.67 	            & 	0.36    &	D 	& 7.08  & 7.50\\			
Alkali-		    &	NaBr		    &  {\em Fm-3m}      &	0.01, 6.17	          	&	0.34    & 	D 	& 6.62  & 7.02\\			
Halides		    &	NaI		        &  {\em Fm-3m}      &	0.01, 5,06             	& 	0.29    & 	D 	& 5.49  & 5.89\\	
			&	RbF		        &  {\em Fm-3m}      &	0.05, 12.93         	& 	0.50    & 	I 	& 10.91 & 10.40\\  
\hline
			&	Na$_2$S	        &  {\em Fm-3m}	    &	0.06, 5.09            	& 	0.31    & 	D 	& 4.22  & -   \\
			&	Na$_2$Se        &  {\em Fm-3m}      & 	0.05, 4,34	         	& 	0.26    & 	D 	& 3.52  & -   \\		
			&	Na$_2$Te	    &  {\em Fm-3m}    	&	0.02, 3,55           	&	0.24 	& 	D 	& 3.21  & -   \\
Alkali-		    &	K$_2$S	        &  {\em Fm-3m}    	&	0.18, 4,70           	& 	0.37    & 	I 	& 4.22  & -   \\		
Chalcogenides	&	K$_2$Se	        &  {\em Fm-3m}  	&	0.18, 4,40           	& 	0.37    & 	I 	& 3.91  & -   \\		
			&	K$_2$Te	        &  {\em Fm-3m}     	&	0.09, 3,73           	& 	0.31    & 	I 	& 3.71  & -   \\
			&	Rb$_2$S	        &  {\em Fm-3m}     	&	0.13, 4.77           	& 	0.36    & 	I 	& 3.82  & -   \\		
			&	Rb$_2$Se        &  {\em Fm-3m}      &	0.12, 3.97	          	& 	0.33    & 	I 	& 3.36  & -   \\	
\hline\hline	
\bf{Ternary}	&	Mg$_2$NF     	& {\em I4$_{1}$/amd}&	0.22, 3.85, 8.33  	    &	0.31	&	I	& 3.17  & -   \\
                &	Cs$_2$SnF$_6$	& {\em P-3m1}		&	0.00, 0.12, 12.27      	&	0.19	&	D	& 8.74  & -   \\
 			&	Rb$_2$CdCl$_4$ 	& {\em I4/mmm}	    &	0.00, 8.55, 6.56      	&	0.43	&	I	& 5.20  & -   \\ 
Halides		    &	Cs$_2$CdCl$_4$ 	& {\em I4/mmm}   	&	0.00, 8.61, 6.67     	&	0.44	&	I	& 5.17  & -   \\
			&	NaHF$_2$ 		& {\em R-3m}		&	0.01, 0.13, 11.97  	    &	0.44	&	I	& 11.88 & -   \\	
                &	Rb$_2$SnBr$_6$	& {\em Fm-3m}  		&	0.00, 0.03, 5.93      	&	0.39	&	D	& 2.19  & -   \\
 			&	Cs$_2$SnBr$_6$ 	& {\em Fm-3m}	    &	0.00, 0.04, 5.97      	&	0.46	&	D	& 2.33  & -   \\    
                &	K$_2$SnBr$_6$	& {\em Fm-3m}		&	0.00, 0.03, 5.90      	&	0.35	&	D	& 2.07  & -   \\
 			&	Ag$_2$HgI$_4$ 	& {\em I4/mmm}	    &	8.57, 7.62, 3.22      	&	0.29	&	D	& 2.11  & -   \\
\hline
Nitrides        &	LiZnN         	& {\em F-43m}		&	1.60, 13.36, 3.05      	&	0.25	&	D	& 0.61  & -   \\
 			%&	CsCdBr         	& {\em Pm3m}	    &	                     	&	0.36   	&	I	&       & -   \\
                &	Li$_4$Na$_2$N$_2$& {\em Fm-3m}		&	0.70, 1.07, 3.10      	&	0.36	&	D	& 1.51  & -   \\
\hline
Chalcogenides   &	NaInSe$_2$    	& {\em R-3m}	    &	0.03, 15.42, 1.25      	&	0.06	&	I	& 2.07  & -   \\
\hline\hline	
\bf{Quaternary} &SnH$_8$N$_2$F$_6$ 	& {\em P-3m1}		& 0.00, 0.02, 0.42, 8.11	& 	0.34	&	D	& 7.18  & -   \\
Halides		    &SnH$_8$N$_2$Cl$_6$	& {\em Fm-3m}		& 0.02, 0.03, 2.03, 6.84	& 	0.49	&	D	& 3.46  & -   \\
	\hline\hline
   \end{tabular}
\end{center}
\label{tableI}
\end{table*}
%

%%%%%%%%%%%%%%%%%%%%%%%%%%%%%%%%%%%%%%%%%%%%%%%
%%%%%%%%%%%%%%%%%%%%%%%%%%%%%%%%%%%%%%%%%%%%%%%
In order to rely on accurate electronic structures for the selected non-oxide materials, 
we have performed  DFT+U calculations by employing the ACBN0 pseudo-hybrid approach for the evaluation of the Hubbard  parameters $U$ \cite{Agapito2015ACBN0}. This has the main advantage of producing accurate values for energy gaps, thus curing the $E_g$ underestimation typical of DFT at a lower computational cost than higher-level theoretical approaches (such as hybrid functionals, or many body GW approaches). 
For each material, the $U$ potentials have been calculated self-consistently for both metals and not-metals elements, the resulting values are reported in Table \ref{tableI}, along with the corresponding band gap. The band structure plot of all compounds are summarized in Figures (S1--S4) of the SI. Notably, the inclusion of the $U$ correction affects neither the effective mass $m^*$ nor the $E_1$ parameters, which keep on satisfying the filtering criteria of Eqs. \ref{eq2}-\ref{eq3}.

%%%%%%%%%%%%%%%%%%%%%%%%%%%%%%%%%%%%%%%%%%%%%%%%%
\section{Magnetic TC discovery} \label{results}
%%%%%%%%%%%%%%%%%%%%%%%%%%%%%%%%%%%%%%%%%%%%%%%%%

In order to induce magnetism in TCs, it is necessary to include dopants into the hosts and evaluate both the electronic and the optical response of the system to assure that doping promotes electrons in the conduction band without altering its transparency. 
We restricted our attention to the binary compounds of Table \ref{tableI}. 
In particular, we focused on a subset of TM-halides (AgF, AgCl, CuCl, CdCl$_2$) and  an alkali-chalcogenide compound (Na$_2$S), that are representatives of different gap type, anion/cation valence, and stoichiometry.

The comparison between AgF and  AgCl provides information on the effect of the host composition, CuCl highlights the role of the gap (direct vs. indirect), and  
CdCl$_2$ and Na$_2$S are non-monovalent crystals with 1:2 stoichiometric ratio. In the  former case the divalent element is the non-metallic one (i.e. Cl), in the latter case it is the metallic one (i.e. Na). 
The different crystal stoichiometry modifies the dopant coordination and thus the effective  charge donation (if any) to the host conduction band.
For all the selected binary hosts, we evaluated the effect of different dopant species, by adding --- one-by-one --- all 30 TMs elements to the system as metal-substitutional defects. 
For each of the 150 TM-host pairs, we performed a set of first-principles simulations that includes the following steps: i) DFT relaxation of the geometry of the defective structure; ii) re-calculation of the $U$ values for both the dopant and the host following the ACBN0 procedure \cite{Agapito2015ACBN0}, as described in Section \ref{methods}; iii) DFT calculation of the total magnetization,  dielectric function and optical transmittance. 
The total magnetization is obtained from ground state electronic structure of the doped system, resulting from spin-unrestricted  DFT calculations. 
The complex dielectric function $\hat{\epsilon}(E)=\epsilon_r+i\epsilon_i$ is evaluated by using a band-to-band single particle approach which conjugates the Drude model for intraband transitions and the Lorentz-model for the interband ones, as a function of the energy ($E$) of the incoming electromagnetic radiation (see Section \ref{methods}).  
Starting from the real $\epsilon_r (E)$ and the imaginary  $\epsilon_i(E)$ part of the dielectric function, the transmittance T($E$) is given by the expression:
\begin{equation} \label{transmittance}
T(E) = 1 - \frac{[1-n(E)]^2+k(E)^2}{[1+n(E)]^2+k(E)^2},
\end{equation}
where 
\begin{eqnarray} \label{n_and_k}
n(E) &=& \sqrt{\frac{\sqrt{\epsilon_r(E)^2+\epsilon_i(E)^2}}{2}+\frac{\epsilon_i(E)}{2} }, \\ \nonumber
k(E) &=& \sqrt{\frac{\sqrt{\epsilon_r(E)^2+\epsilon_i(E)^2}}{2}-\frac{\epsilon_i(E)}{2} }
\end{eqnarray}
are the real and imaginary parts of the complex refraction index $\hat{n}(E)=n(E)+ik(E)$.

In order to be considered as magnetic TCs (MTCs), the resulting dopant-host systems must satisfy simultaneously three conditions: $\mathcal{C}_1$)  the Fermi level 
lies above the top of the first conduction band, so $\Delta E>0$;
$\mathcal{C}_2$) the transmittance is close to unity ($T(E)> 0.9$) in the visible range that corresponds to have a transparent system;
$\mathcal{C}_3$) the total magnetization $\mu$ is greater than zero.
%%%%%%%%%%%%%%%%%%%%%%%%%%%%%%%%%%%%%%%%%%%%%%%
%%%%%%%%%%%%%%%%%%%%%%%%%%%%%%%%%%%%%%%%%%%%%%%
%
\begin{figure}[htb]
\begin{center}
    \includegraphics[width=0.95\columnwidth]{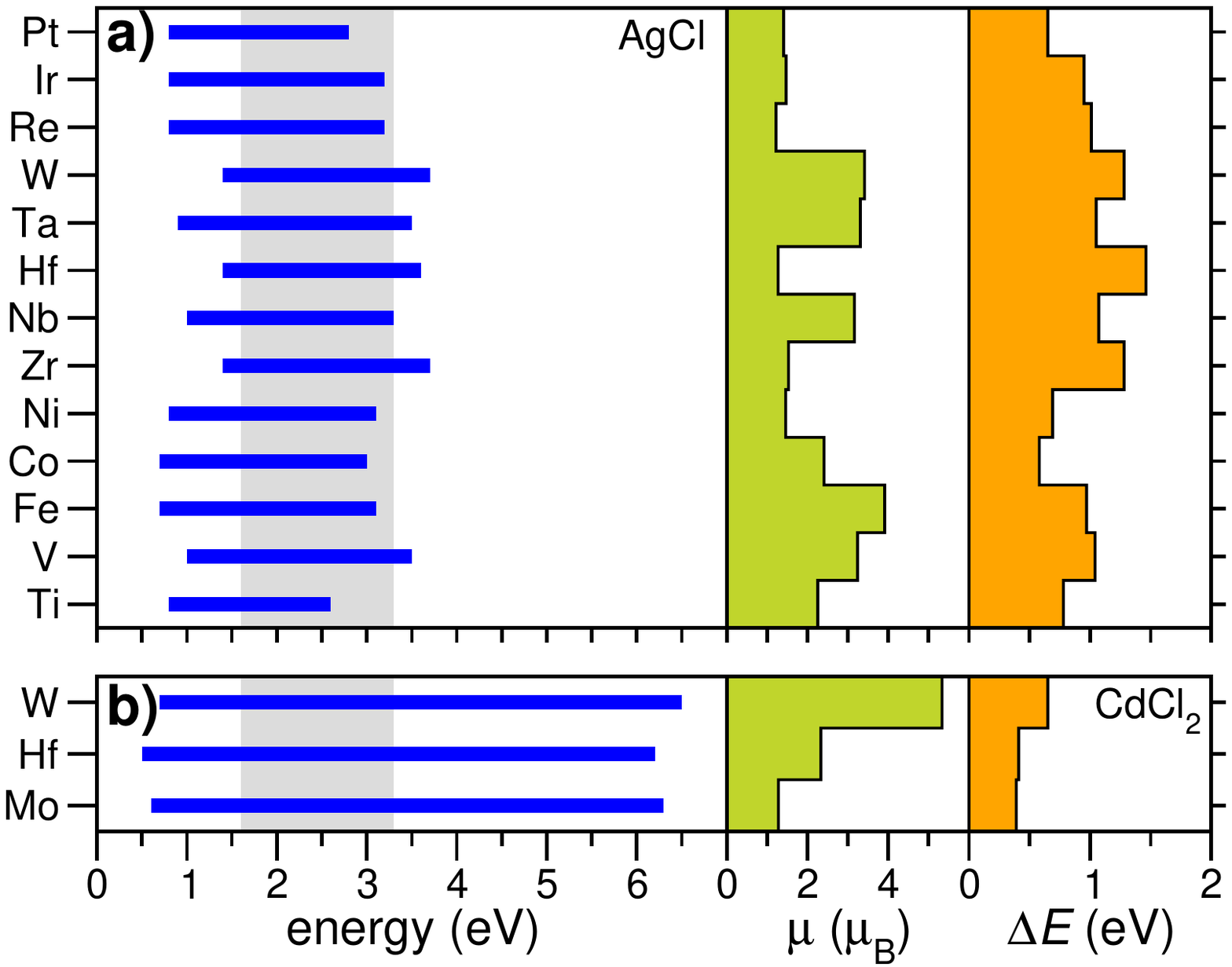}\\ 
    \caption{\small Energy distribution of the optical transmittance $T(E)$ (left panel), total magnetization $\mu$ (central panel),  and  $\Delta E$ (right panel) for (a) TM:AgCl and  (b) TM:CdCl$_2$ magnetic TCs. 
    Blue horizontal lines identify the transparency range $T(E)>0.9$; the shaded gray area indicates the visible range. Only the TM-host systems that satisfy the three conditions $\{\mathcal{C}_i\}$ are reported.}
    \label{AgCl-CdCl2}
  \end{center}
\end{figure}
%

%%%%%%%%%%%%%%%%%%%%%%%%%%%%%%%%%%%%%%%%%%%%%%%
%%%%%%%%%%%%%%%%%%%%%%%%%%%%%%%%%%%%%%%%%%%%%%%
All five representative host candidates are suitable for the design of magnetic TC with several TM doping elements.  53 host-dopant combinations over the 150 possible compositions fulfill the filtering requirements.
The results of the HT search are summarized in Figure \ref{AgCl-CdCl2} for (a) AgCl and (b) CdCl$_2$ hosts, respectively. Frequency regions where $T(E)>0.9$ (blue bars) and the visible range (gray area) are shown together with total magnetization and $\Delta E$.

Note that not all TMs are optimal dopants to have magnetic TCs: the number and the chemical elements vary with the host. In the case of AgCl, 13 of the 30 dopants satisfy the criteria, whereas only 3 TMs give a positive outcome in the case of CdCl$_2$. In TM:AgCl the transparency windows mostly cover the visible range, with minor extensions into the near-IR  and/or near-UV. Conversely, in TM:CdCl$_2$ systems the optical transparency range broadly  spans the entire range from mid-IR to deep-UV (0.5-6.5 eV). This different behavior is related to the different bandgap of the pristine host semiconductors (see Table \ref{tableI}).
Results for AgF, CuCl and Na$_2$S are reported in Figures S5-S7 of SI. In particular, AgF and Na$_2$S are similar to AgCl for what concerns the optical properties, having a transparency window that covers almost entirely the visible spectrum.
The case of CuCl is different since the transparency range lies between the IR and visible range (Figure S6, SI). 

\begin{figure}[htb]
\begin{center}
    \includegraphics[width=0.95\columnwidth]{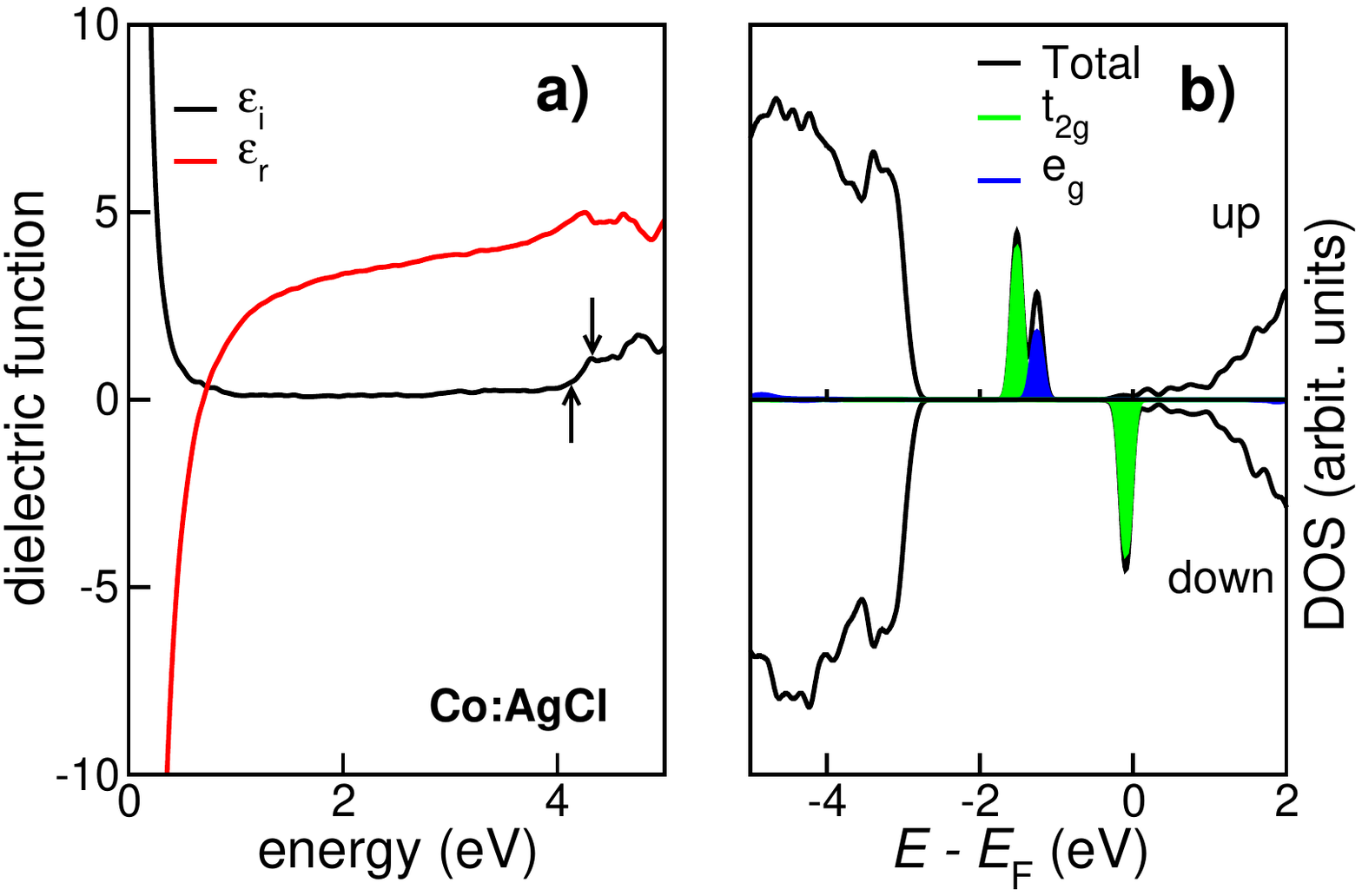}
    \caption{\small (a) Real and imaginary parts of the dielectric function  and (b) spin-projected  DOS  for Co:AgCl. Zero energy reference in panel (b) is fixed at the Fermi level of the doped system. In panel (a) the two black arrows indicate the energy at which the vertical transitions start to occur for pristine AgCl (lowest energy, up arrow) and Co:AgCl (highest energy, down arrow). In panel (b) the spin up and spin down projections of the DOS are reported together with the projection on the $t_{2g}$ and the $e_g$ orbitals that Co forms hybridizing with the host cubic crystal.} 
    \label{Co:AgCl}
  \end{center}
\end{figure}

The number of host-dopant combinations giving a positive outcome as MTCs is 13, 12 and 12 for AgF, CuCl and Na$_2$S respectively.
Figure \ref{Co:AgCl} reports the dielectric function and the density of states (DOS) for the case of Co:AgCl, assumed as prototypical MTC.
%%%%%%%%%%%%%%%%%%%%%%%%%%%%%%%%%%%%%%%%%%%%%%%
%%%%%%%%%%%%%%%%%%%%%%%%%%%%%%%%%%%%%%%%%%%%%%%
%
%

%%%%%%%%%%%%%%%%%%%%%%%%%%%%%%%%%%%%%%%%%%%%%%%
%%%%%%%%%%%%%%%%%%%%%%%%%%%%%%%%%%%%%%%%%%%%%%%
Doping is responsible for the Drude-like shape of the dielectric function typical of metals (Figure \ref{Co:AgCl}-a), with the real ($\epsilon_r$, red curve) and imaginary ($\epsilon_i$, black line) parts diverging for $E\rightarrow 0$. This behavior is the fingerprint of charge carrier presence in the conduction band and indicates that a insulator-to-conductor transition has occurred. 
Nonetheless, the imaginary part, which accounts for the optical absorption, is close to zero in the entire visible range (1.6--3.3 eV). This corresponds to transparency.
We note also that the interband transition threshold, which is representative of the bandgap, is blue shifted with respect to the undoped host, in agreement with the Burnstein-Moss model: this confirms the shift of the Fermi level in the conduction band and the formation of a degenerate n-type semiconductor. 

The presence of Co induces into the system  a total magnetization ($\mu=2.24~\mu_0/$cell), which derives from the imbalance between the occupation of the spin-up and spin-down states, as shown in Fig.\ref{Co:AgCl}-b. 
The spin-up spectrum has two sharp peaks within the host bandgap. These correspond to the $t_{2g}$ and $e_g$ orbitals that Co forms within the cubic crystal field in the host (green and blu areas, panel b). Both states are fully occupied and may be optically active. However, their energy position in the bandgap is deep enough to avoid  perturbing  the transparency of the system (Figure \ref{AgCl-CdCl2}-a). In the case of  spin-down, only the  $t_{2g}$ peak is  occupied and degenerate with the Fermi level (zero energy reference in Figure \ref{Co:AgCl}-b). After doping the system is electrically conductive, optically transparent, and it has a net magnetic moment, i.e. it is a {magnetic transparent conductive} compound.

Similar properties are shared by all the 53 dopant-host combinations resulting in MTCs, although the dielectric functions as well as the details of the DOSs (e.g. number and symmetry of occupied states, energy position of $E_F$) 
depend on the specific chemistry and cause the differences observed in Figure \ref{AgCl-CdCl2}. By comparing the results across the full HT study,  we 
extracted the following information {\bf (i)} the type of host bandgap (direct vs. indirect) is irrelevant for the realization of MTCs; {\bf (ii)} the same conclusion holds for isoelectronic compositions of the hosts (e.g. AgF and AgCl); {\bf (iii)} the band alignment between the dopant ionization potential and the host bandgap affects the position of the metal-derived states within the bandgap. The higher is the bandgap the lower is the possibility for the dopant electrons to reach the  conduction band and inject free charge into the host. This explains why only three TM:CdCl$_2$ pairs satisfy the filtering conditions $\{\mathcal{C}_i\}$, being the remaining systems non-conductive (i.e. localized in-gap defect states). 
{\bf (iv)} The transparency in a given energy range is not directly related to the spin polarization of the doped system, and
{\bf (v)} even though the magnetic character of the dopant elements imparts a spin-polarization to the host, the localized character of the resulting states prevents the establishment of an electron-itinerant magnetism in the MTCs (e.g. extended ferromagnetic or antiferromagnetic behavior). Although all resulting MTCs can potentially act as diluted magnetic semiconductors \cite{Zunger10}, the presence of spatially localized magnetic moments does not allow for the emergence of itinerant ferromagnetic states. On the contrary, the localization of the magnetic d-states of the dopands suggests that those system can rather be exploited for spin-filtering in spintronic applications.

A spin-filter allows electrons with a fixed spin polarization (e.g. spin up) to pass, and blocks electrons with opposite polarization (e.g. spin down). This implies a sizable difference between the spin-polarized electron conductivities.
Spin-polarization is a necessary  but not-sufficient condition for a spin-polarized conductivity: 
what is crucial 
%role 
for the electron transport is
the energy alignment among $E_F$, the spin-polarized orbitals, and the conduction band. 
The first requirement for conductivity is that
$E_F$ must lie in the conduction band of the host. This condition is always fulfilled  (by definition) in TCs. Depending on the energy position of the Fermi level with respect to the spin-unpaired orbitals
two different scenarios are possible. In the first case, the orbitals are degenerate with $E_F$ and there is a spin-imbalance in the conducting electrons resulting in a difference in the spin conductivity (spin filter). In the second case, the electronic states associated with the defect lie energetically far away from $E_F$ and the conducting properties of the system will be  spin-independent (no spin filter).
In principle, by applying external voltages it is always possible to align the Fermi level to the energy of a spin-polarized peak. However, here we focus only on the intrinsic properties of the doped systems (i.e. low voltage regime $E_F\pm0.5$ eV), as the application of strong  external electric field could be detrimental for the transparency of the systems, 
 hence destroying the TC character of the compound.
%%%%%%%%%%%%%%%%%%%%%%%%%%%%%%%%%%%%%%%%%%%%%%%
%%%%%%%%%%%%%%%%%%%%%%%%%%%%%%%%%%%%%%%%%%%%%%%
\begin{figure}[htb]
\begin{center}
    \includegraphics[width=0.95\columnwidth]{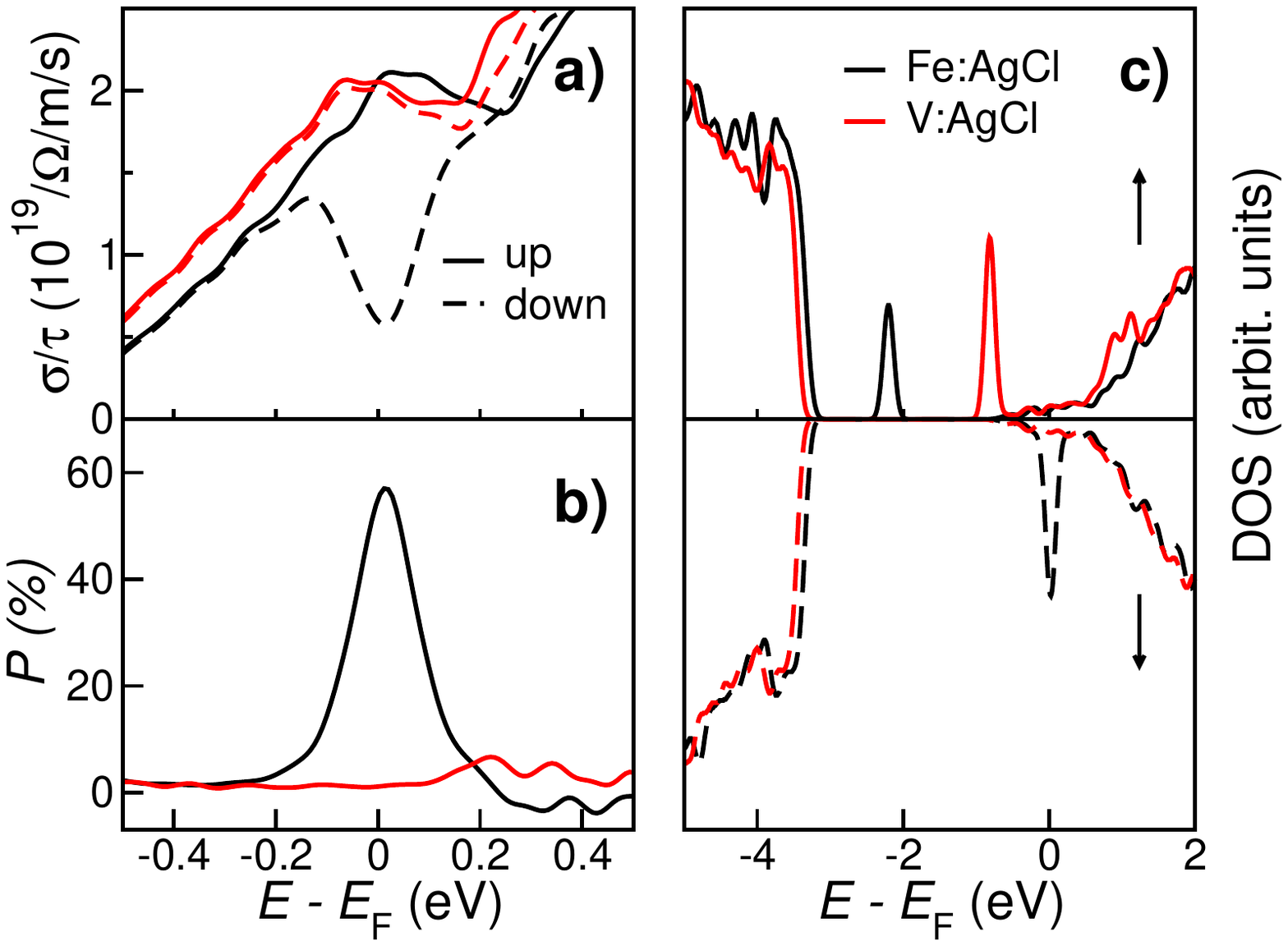}
    \caption{\small (a) Conductivity, (b) spin-polarization, P, and (c) DOS for V:AgCl (black curves) and Fe:AgCl (red curves). Spin-up and spin-down components are represented by solid and dashed lines respectively. For Fe:AgCl there is a spin-down polarized state centered at the Fermi level, that corresponds to an enhancement of the spin-up conductivity and resulting in pronounced polarization. On the contrary for V:AgCl there are no spin-polarized states at the Fermi level, resulting in an almost spin-independent conductivity and to a polarization close to zero.}
    \label{V-Fe:AgCl}
  \end{center}
\end{figure}
%

%%%%%%%%%%%%%%%%%%%%%%%%%%%%%%%%%%%%%%%%%%%%%%%
%%%%%%%%%%%%%%%%%%%%%%%%%%%%%%%%%%%%%%%%%%%%%%%
We compared  two cases (namely V:AgCl and Fe:AgCl) corresponding to the two  scenarios described above.  V:AgCl and Fe:AgCl are both MTC compounds,  but have rather different electrical conductivity. 
Figure \ref{V-Fe:AgCl}-a shows the spin-polarized conductivity ($\sigma^{\uparrow}$,$\sigma^{\downarrow}$)  for spin up ($\uparrow$) and spin down ($\downarrow$) channels calculated within the constant relaxation time approximation ($\sigma/\tau$) of the Boltzmann's transport equation \cite{damico2016}, within the pseudo-atomic orbital scheme implemented in the PAOFLOW code \cite{paoflow} (see Section \ref{methods}). 
The differences between the two systems are evident: while V:AgCl has similar conduction properties for both spin up and spin down with a relative maximum at $E=E_F$, Fe:AgCl exhibits a different behavior at the Fermi level, where $\sigma^{\uparrow}$ has a maximum and $\sigma^{\downarrow}$ has a minimum. In order to quantify this difference we define the {spin conduction polarization}:

\begin{equation}
P = \frac{\sigma^{\uparrow}-\sigma^{\downarrow}}{\sigma^{\uparrow}+\sigma^{\downarrow}},
\label{spin-pol}
\end{equation}
whose plots for V:AgCl and Fe:AgCl are reported in Figure \ref{V-Fe:AgCl}-b.
It turns out that, even though both systems have a net magnetic moment (i.e. spin-imbalance), V:AgCl is almost insensitive  to the spin polarization of the transported electrons, while Fe:AgCl acts as a good spin filter, blocking more than  $\simeq$ 60\% of spin down electrons. The microscopic origin of this difference relays on the relative energy position of the defect states and of the Fermi energy (panel c). In the case of iron (black lines), the spin up conductivity for $E=E_F$ derives from the almost parabolic bottom of the AgCl conduction band. The $e_{g}$ defect state is fully occupied at $\sim 2$ eV below the Fermi level and does not contribute to transport. In the spin down spectrum, the $t_{2g}$ is degenerate with the bottom of the AgCl conduction band and aligned with the position of the Fermi level. This orbital is spatially strongly localized and operates as a sink for spin down electrons, giving rise to a net spin conduction polarization (black peak in panel b).     
In the case of V:AgCl (red lines), there are no V-derived states aligned with E$_F$ and the conductivity is due for both spins to the conduction band of the host, which is, indeed, very similar to  $\sigma^{\uparrow}$ component of Fe:AgCl. This corresponds to a vanishing value of $P$.
Intermediate configurations, where a defect peaks are  partially centered on (or in proximity of) $E_F$, will give rise to intermediate values of $P$. 

Starting from the MTCs  resulting from our HT research,  we investigated which dopant-host combinations are also suitable for spin-filter applications.  We calculated $P$ for all the MTCs, the results are summarized in Table \ref{polarization}, where we reported the maximum value of $P$ (expressed in percentage) in the energy window [-0.5; +0.5] eV around the Fermi level. 
$P$ spans from positive (spin-up) to negative (spin-down) values, depending on the presence of states localized on the dopant across the Fermi energy and on its spin. $P$ can reach very large values, for example  +90$\%$ for W:CdCl$_2$ and  -70$\%$ for Zr:AgCl, confirming that several TCMs may act as good spin filters. More specifically, by fixing a threshold of $\pm50\%$ to consider a host-dopant system suitable as a transparent conductor spin-filter, we obtain 15 possible candidates (highlighted in Table \ref{polarization} with a x ), 5 of those with $P$ larger than 70$\%$ ({\bf X} ).
%%%%%%%%%%%%%%%%%%%%%%%%%%%%%%%%%%%%%%%%%%%%%%%
%%%%%%%%%%%%%%%%%%%%%%%%%%%%%%%%%%%%%%%%%%%%%%%
%
\begin{table*}[htb]
\caption {Maximum value of the spin conduction polarization $P$ within the energy window [-0.5; +0.5] eV around the Fermi level for all the systems fulfilling the $\mathcal{C}_i$ criteria. The percentages of polarization are reported for each host-dopant combination, defining if the system can act as a good spin-filter.}
   \begin{center}   
    \begin{tabular}{c||c|c|c||||c||c|c|c}
     \hline\hline
 			  & Dopant	 & Polarization ($\%$)& Spin-filter	& & Dopant	 & Polarization & Spin-filter	\\
    \hline\hline
%\bf{Binary} & & & & & & & \\ 
         	&	 V		&  -39          &	        &               	&	V		&  +11          &	     \\			
			&	Cr		&  +3          	&		    &                   &	Fe		&  +57        	&	x \\			
			&	Mn  	&  -36        	&		    &                   &	Co  	&  +21       	&	     \\	
			&	Fe		&  +50          &	x	    &                   &	Ni		&  +35          &	     \\			
		&	Co		&  +40          &	        &                   &	Zr		&  -70          &	\bf{X}  \\
		    &	Ni   	&  +1          	&	        &                   &	Nb   	&  -57       	&	x  \\		
        AgF    	&	Mo     	&  -38          &	        &       AgCl        &	Hf     	&  -43          &	     \\		
			&	Tc     	&  +48      	&	        &                   &	Ta     	&  -58         	&	x  \\		
			&	Pd     	&  +38      	&	        &                   &	W     	&  +57      	&	x  \\		
			&	W       &  -47         	&	        &                   &	Re      &  +61         	&   x  \\
   			&	Re     	&  +15       	&	        &                   &	Ir     	&  -49         	&	     \\		
			&	Os     	&  -3       	&	        &                   &	Pt     	&  +43         	&	     \\		
			&	Pt      &  -32         	&	        &                   &  	        &           	&	     \\
\hline\hline	
        	&	Ti		&  +6           &	 	    &               	&	Ti		&  +1           &	     \\			
			&	V		&  +8       	&		    &                   &	Mn		&  -43        	&	     \\			
			&	Cr  	&  +39        	&		    &                   &	Fe  	&  0         	&	     \\	
			&	Mn		&  +29          &	        &                   &	Co		&  +70          &	\bf{X}  \\			
	    	&	Co		&  +62          &	x     &                   &	Zr		&  +1           &	     \\
    CuCl 	    &	Zr   	&  +10         	&	        &       Na$_2$S     &	Nb   	&  +1       	&	     \\		
			&	Tc     	&  +71          &	\bf{X}     &                   &	Tc     	&  0            &	     \\		
			&	Hf     	&  +11      	&	        &                   &	Hf     	&  +6          	&	     \\		
			&	Ta     	&  0        	&	        &                   &	Ta     	&  +1          	&	     \\		
			&	W       &  +61         	&	x     &                   &	Re      &  0        	&        \\
   			&	Re     	&  0         	&	        &                   &	Ir     	&  +65         	&	x  \\		
\hline\hline	
        	&	Mo		&  +59          &	x	   \\			
	    CdCl$_2$	&	Hf		&  -80      	&	\bf{X}	   \\			
			&	W   	&  +92        	&	\bf{X}	   \\	
	\hline\hline
   \end{tabular}
\end{center}
\label{polarization}
\end{table*}
%%%%%%%%%%%%%%%%%%%%%%%%%%%%%%%%%%%%%%%%%%%%%%%
%%%%%%%%%%%%%%%%%%%%%%%%%%%%%%%%%%%%%%%%%%%%%%%

%%%%%%%%%%%%%%%%%%%%%%%%%%%%%%%%%%%
\section{Computational details} \label{methods}
%%%%%%%%%%%%%%%%%%%%%%%%%%%%%%%%%%%

First principles calculations have been performed within local-spin approximation and the DFT+U framework, by using the Quantum Espresso suite \cite{qe}.  The ACBN0 approach \cite{Agapito2015ACBN0}  has been used to evaluate the Hubbard  $U$ parameters for each chemical species and for all dopant-host pairs.
We first calculated the values of the $U$ parameters for the host materials, and we used them 
 to obtain the electronic structure of the undoped materials (see Table \ref{tableI}). Next, by keeping fixed the $U$ values for the host, we calculated the corresponding values for the dopant within the host and we refined the electronic structure of the doped system.
The calculations for the $U$ parameters for the hosts have been performed using norm-conserving PBE pseudopotentials with a ($4 \times 4 \times 4$) grid of k-points to sample the Brillouin zone,  a kinetic-energy cut-off of 150 Ry
for the plane-wave expansion of the single particle wavefunction.

We have then simulated ($2\times2\times2$) cubic supercells containing 64 atoms for AgF, AgCl and CuCl and 81 atoms for Na$_2$S and CdCl$_2$, using one substitutional atom per cell resulting in a doping percentage of 1.56\% or 1.12\% for the two  cases. 

The optical properties have been calculated with the {\em epsilon.x} code, also included in the Quantum Espresso suite. This code evaluates the complex dielectric function  $\hat{\epsilon}(\omega)$ within the Drude-Lorentz approximation as described by the following relations  \cite{benassi2007}:
%%%%%%%%%%%%%%%%%%%%%%%%%%%%%%%%%%%%%%%%%%%%%%%
%%%%%%%%%%%%%%%%%%%%%%%%%%%%%%%%%%%%%%%%%%%%%%%
\begin{equation}
\hat{\epsilon}(\omega)= 1 - \sum_{{\bf k},n}f_{{\bf
k}}^{n,n}\frac{\omega_p^2}{\omega^2+i\eta\omega} + \sum_{{\bf
k},n\neq n'}f_{{\bf k}}^{n,n'}\frac{\omega_p^2}{\omega^2_{{\bf
k},n,n'}-\omega^2-i\Gamma\omega}, \label{eps}
\end{equation}
where
\begin{equation}
\omega_{p}=\sqrt{\frac{4\pi e^2N_e}{m}} \label{plasma}
\end{equation}
%%%%%%%%%%%%%%%%%%%%%%%%%%%%%%%%%%%%%%%%%%%%%%%
%%%%%%%%%%%%%%%%%%%%%%%%%%%%%%%%%%%%%%%%%%%%%%%
is the bulk plasma frequency; $\hbar\omega_{{\bf k},n,n'}= E_{{\bf k},n}-E_{{\bf
k},n'}$ is the vertical band-to-band transition energy between
occupied  and empty  Bloch states labeled by the quantum numbers
$\{{\bf k},n\}$  and $\{{\bf k},n'\}$.  The Drude-like and Lorentz-like relaxation terms, $\eta, \Gamma \rightarrow
0^+$, are
associated to intra-band and inter-band transitions respectively; $f_{{\bf k}}^{n,n}$ and $f_{{\bf k}}^{n,n'}$ are the
corresponding oscillator strengths.

The electrical spin conductivity  $\sigma^{s}$ ($s=\{\uparrow,\downarrow\}$) is evaluated by 
using the PAOFLOW code \cite{paoflow} that solves the Boltzmann equation for transport in the scattering-time approximation, as \cite{grosso2000}:
\begin{equation}
\sigma^{s}=\frac{e^2}{4\pi^3}\int_{BZ}\tau \sum_n v_n^s(\boldsymbol{k})v_n^s(\boldsymbol{k})\Big(-\frac{ \partial f_0 (T) }{ \partial \mathcal {E} } \Big ) d\boldsymbol{k},
\label{eq:cond}
\end{equation}
where $\tau$ is the constant relaxation time, $v_n^s(\boldsymbol{k})$ is the electron velocity $\hat{v}$ calculated for the $n^{th}$ spin-polarized ($s$) band for each $\boldsymbol{k}$ point in the BZ, $f_0(T)$ is the equilibrium distribution function at the temperature $T$, and $\mathcal {E}$ is the electron energy.
The evaluation of Eq. \ref{eq:cond} requires an accurate integration over a fine grid of $k$ point in the BZ. Here, we used a tight-binding like representation obtained from a pseudo-atomic orbital (PAO) projection procedure of the DFT electronic wavefunctions originally expressed in plane waves\cite{paoflow}.
The group velocity is calculated from the expectation value of the momentum operator $\hat{p}$ \cite{damico2016}: 
\begin{eqnarray}
    v_n^s(\boldsymbol{k})&=& \frac{1}{\hbar}\langle \psi_n^s(\boldsymbol{k})|\frac{\hat{p}}{m_0}|\psi_m^s(\boldsymbol{k})\rangle \\ 
    &=& \frac{1}{\hbar}\langle u_n^s(\boldsymbol{k})|\nabla_{\boldsymbol{k}}\hat{H}(\boldsymbol{k})|u_m^s(\boldsymbol{k})\rangle,\nonumber
    \label{eq:vel}
\end{eqnarray}
where $m_0$ is the free electron mass and $|\psi_n^s(\boldsymbol{k})\rangle=exp(-i\boldsymbol{k}\cdot\boldsymbol{r})|u_n^s(\boldsymbol{k})\rangle$ are the Bloch functions resulting from the spin-unrestricted DFT calculations. The gradient of the Hamiltonian operator that enters in Eq. \ref{eq:vel} is easily evaluated in terms of the TB hamiltonian $\hat{H}(\boldsymbol{r}_{\ell})$ obtained projecting the Bloch wavefunctions onto a PAO basis set \cite{curtarolo:art86}: 
\begin{equation}
    \nabla_{\boldsymbol{k}}\hat{H}(\boldsymbol{k})=\sum_{\ell}i\boldsymbol{r}_{\ell}exp(-i\boldsymbol{k}\cdot\boldsymbol{r}_{\ell})\hat{H}(\boldsymbol{r}_{\ell}).
    \label{eq:ham}
\end{equation}
Here,  spin conductivities for the doped systems have been calculated with a dense grid of ($20 \times 20 \times 20$) k-points. 
%%%%%%%%%%%%%%%%%%%%%%%%%%%%%%%%%%%%%%%%%%%%%%%%%%%%%%%%%%%%%
\section{Conclusions} \label {conclusions}
%%%%%%%%%%%%%%%%%%%%%%%%%%%%%%%%%%%%%%%%%%%%%%%%%%%%%%%%%%%%%

Using first principles high-throughput approaches, we theoretically predicted a novel class of non-oxide materials, namely magnetic transparent conductors, that merge the optical and transport properties typical of TCs and a net magnetic moment.
We first identify  possible hosts that can be used to realize MTC compounds by doping with transition metal element. Then, we characterized the electronic and optical properties of the resulting MTC systems. 
Finally, our approach clearly identified a set of MTCs
that exhibit  a large spin conduction polarization up to $\simeq$ 90\% and that  can be exploited for spin filtering. The discovery of  this  new class of magnetic transparent conductors may open new routes for application of TC compounds, for example as spin filters in spintronic
devices.

\section{Acknoledgments}

A.R. acknowledges the project funded under the National Recovery and Resilience Plan (NRRP), Mission
04 Component 2 Investment 1.5, NextGenerationEU, Award Number:000105.
Ar.C. acknowledges the 
National Centre for HPC, Big Data and Quantum Computing (ICSC), 
funded under the National Recovery and Resilience Plan (NRRP), Mission
04 Component 2 Investment 1.4, NextGenerationEU, Award Number:CN00000013.
%%%%%%%%%%%%%%%%%%%%%%%%%%%%%%%%%%%%%%%%%%%%%%%
%%%%%%%%%%%%%%%%%%%%%%%%%%%%%%%%%%%%%%%%%%%%%%%
\medskip
\newcommand{\Ozolins}{Ozoli\c{n}\v{s}}

\end{document}

\pagestyle{fancy}
%\rhead{\includegraphics[width=2.5cm]{vch-logo.png}}

\title{Supporting Information for: Magnetic Transparent Conductors for Spintronic Applications}

\maketitle

\author{Pino D'Amico*,} 
\author{Alessandra Catellani,}
\author{Alice Ruini,}
\author{Stefano Curtarolo,}
\author{Marco Fornari,}
\author{Marco Buongiorno Nardelli,} 
\author{Arrigo Calzolari**}

\begin{affiliations}
P. D'Amico*, A. Catellani, A. Calzolari** \\
Istituto Nanoscienze CNR-NANO-S3, I-4115 Modena, Italy\\
*Email: pino.damico@nano.cnr.it
**Email: arrigo.calzolari@nano.cnr.it

A. Ruini \\
Dipartimento di Fisica, Informatica e
Matematica, Universit\'a di Modena e Reggio Emilia, I-41125
Modena, Italy

S. Curtarolo\\
Center for Materials Genomics, Duke University, Durham, NC 27708, USA\\
Materials Science, Electrical Engineering, Physics and Chemistry, Duke University, Durham, NC 27708, USA

M. Fornari\\
Department of Physics, Central Michigan University, Mt. Pleasant, MI 48859\\
Center for Materials Genomics, Duke University, Durham, NC 27708, USA

M. Buongiorno Nardelli\\
Department of Physics, University of North Texas, Denton, TX 76203, USA\\
Center for Materials Genomics, Duke University, Durham, NC 27708, USA\\
\end{affiliations}

%%%%%%%%%%%%%%%%%%%%%%%%%%%%%%%%%%%%
\section{Full list of filtered TCs}
%%%%%%%%%%%%%%%%%%%%%%%%%%%%%%%%%%%%

In the following we report the complete list of the 115 materials resulting from the screening of the {\em AFLOWLIB} database, according to the filtering criterions described in the main text.
The materials are listed in separate tables for binary (Table \ref{tab:binary} and Table \ref{tab:binary_continued}), ternary (Table \ref{tab:ternary}) and quaternary (Table \ref{tab:quaternary}) compounds and for each entry we report the crystal structure, the ICSD number (useful for a quick search on the {\em ALFOWLIB} online web interface at https://www.aflowlib.org/), the effective mass of the first conduction band, the gap type and the value of the energy gap as present in the database.
\newpage
\begin{table*}[htb]
\caption {Binary compounds resulting from screenig the AFLOW database with the descriptors in Eqs.~(1-3) of the main text. The list contains information on  the symmetry group,  effective mass ($m^*/m_0$), gap type and its value $E_g^{Th}$ as founded in the {\em AFLOWLIB} repository.}
%\caption {Screened  binary compounds.}
    \begin{center}
        {\bf{BINARY COMPOUNDS}}\\
        \vspace{3mm}
        \begin{tabular}{ c|c|c|c|c|c|c  }
    \hline\hline
                &Structure & Material & ICSD number & $m^*/m_0$ & Gap  & $E_g^{Th}(eV)$ \\
    \hline\hline
    \bf{Oxides} & RHL    & Al$_2$O$_3$    &  30026    & 0.440 & D & 5.9155 \\
                & MCLC   & Ga$_2$O$_3$    &  184327   & 0.224 & I & 2.1538 \\
                & BCC    & In$_2$O$_3$    &  14387    & 0.274 &	D & 1.1134 \\ 	
                & HEX	 & ZnO            &  180052   &	0.406 &	D & 1.8238  \\
                & FCC	 & Na$_2$O        &  644917	  & 0.355 & I & 1.9251 \\ 		
                & FCC	 & K$_2$O         &  180571   & 0.403 & I & 1.7185\\		
                & FCC	 & Rb$_2$O        &  77676 	  & 0.357 & I & 1.3099 \\		
                & ORC	 & GeO$_2$        &  281600   & 0.422 & D & 1.4591 \\
    \hline\hline
    \bf{Non-Oxides} & HEX	 & GaN    & 181358    & 0.339 & D & 1.9259  \\
                    & FCC	 & YN     & 183180    & 0.338 & I & 2.1202  \\
                    & FCC	 & CdS    &  31075    &	0.189 &	D & 1.2648  \\	
                    & FCC	 & AgF &  18008   &	0.412 &	I & 1.1242  \\			
                    & FCC      & AgCl &  56539  &	0.256 &	I & 1.3969  \\			
                    & FCC	 & AgBr &  157536 &	0.225 &	I & 1.6258  \\			
                    & FCC	 & CuCl &  60711  &	0.325 &	D & 1.3275  \\			
                    & FCC	 & CuBr &  78275  &	0.243 &	D & 1.2007  \\
                    & RHL	& ZnBr$_2$ &  26080  & 0.123 & I & 3.5257 \\		
                    & FCC     & CdF$_2$ &  28864   & 0.406 & I & 3.0578 \\		
                    & RHL	& CdCl$_2$ &  62202  & 0.207 & I & 3.5881 \\		
                    & RHL	& CdBr$_2$ &  52367  & 0.152 & I & 2.8620 \\	
                    & FCC	& HgF$_2$ &  33614   & 0.308 & D & 1.0340 \\
                    & FCC	& LiCl    &  52235 	& 0.478 & D & 6.2951 \\
                    & FCC	& LiBr    &  44274 	& 0.363 & D & 4.9605 \\			
                    & FCC	& NaBr  &  41440 	& 0.337 & D & 4.1571 \\			
                    & FCC	& NaI     &  44279 	& 0.289 & D & 3.6206 \\	
                    & FCC	& RbF    &  53828 	& 0.499 & I & 5.5613\\ 
                    & FCC	& Na$_2$S   &  644959 	& 0.311 & D & 2.4889 \\
                    & FCC	& Na$_2$Se &  645027 & 0.257 & D & 2.0634 \\		
                    & FCC	& Na$_2$Te &  76553 	& 0.236 & D & 2.0664 \\
                    & FCC	& K$_2$S     &  183837 	& 0.372 & I & 2.3288 \\		
                    & FCC	& K$_2$Se   &  168448 	& 0.336 & I & 2.0882 \\		
                    & FCC	& K$_2$Te    &  182742 	& 0.312 & I & 2.1323 \\
                    & FCC	& Rb$_2$S   &  29208 	& 0.355 & I & 1.9677 \\		
                    & FCC	& Rb$_2$Se &  168449	& 0.327 & I & 1.8007 \\	
        \hline\hline	
   \end{tabular}
    \end{center}
   \label{tab:binary}
\end{table*}

\begin{table*}
\caption {Ternary compounds resulting from screenig the AFLOW database with the descriptors in Eqs.~(1-3) of the main text. The list contains information on  the symmetry group,  effective mass ($m^*/m_0$), gap type and its value $E_g^{Th}$ as founded in the {\em AFLOWLIB} repository.}
\begin{center}
        {\bf{TERNARY COMPOUNDS}}\\
        \vspace{3mm}
    \begin{tabular}{ c|c|c|c|c|c|c  }
    \hline\hline
        & Structure & Material & ICSD & $m^*/m_0$ & Gap  & $E_g (eV)$ \\
    \hline\hline
$\bf{Oxides}$ &  ORCF&	InFO &  2521  &	0.332	&	D & 1.5030  \\
& BCT&	LiInO$_2$ & 639886 & 	0.310	&	I&2.0132 \\
& RHL&	NaInO$_2$ & 34600  &	0.130	&	I&2.0903 \\
& RHL&	LiAlO$_2$ & 28288  &	0.214	&	D&6.1314 \\
& RHL&	KInO$_2$ & 380401  &	0.119	&	I&2.2237 \\
& RHL&	NaAlO$_2$ & 22216  &	0.160	&	D&4.7879 \\
& RHL&	LiGaO$_2$ & 28388  &	0.181	&	I&3.7575 \\
& RHL&	CuGaO$_2$ & 60846  &	0.183	&	I&1.0539 \\
& RHL&	CuAlO$_2$ & 32630  &	0.351	&	I&2.0266 \\
& HEX&	CdH$_2$O$_2$& 165225  &	0.452	&	I&1.8268 \\
& RHL&	CaHgO$_2$ & 80717  &	0.422	&	I&2.1253 \\
& RHL&	ScAgO$_2$ & 422442 & 	0.330	&	I&2.4606 \\
& FCC&	KAlO$_2$ & 262975  &	0.444	&	I&2.9342 \\
& FCC&	RbAlO$_2$ & 28373  &	0.394	&	I&3.3707 \\
& HEX&	Mg$_2$H$_2$O$_3$    & 95472  &	0.234	&	D&3.5675 \\
& ORCC&	Na$_2$SiO$_3$            & 74640  &	0.463	&	I&3.9663 \\
& ORCC&	Na$_2$GeO$_3$          & 1622  &	0.443	&	I&3.0212 \\
& CUB&	CaSnO$_3$                  & 29149  &	0.431	&	I&1.3672 \\
& RHL&	NaSb1O$_3$ 		    & 78416  &	0.381	&	I&2.6830  \\
& RHL&	KSbO$_3$                    & 33546  &	0.349	&	I&2.6805 \\
& RHL&	BaSiO$_3$                    & 156705 & 	0.286	&	I&3.0701 \\
& RHL&	MgGeO$_3$                  & 171787 & 	0.379	&	D&3.6887 \\
& RHL&	ZnSiO$_3$                    & 167186 & 	0.399	&	D&4.4876 \\
& RHL&	MgSiO$_3$                   & 89805  &	0.427	&	D&5.8345 \\
& RHL&	CdGeO$_3$                  & 30971  &	0.218	&	I&1.5976 \\
& RHL&	ZnGeO$_3$                   & 33722  &	0.259	&	D&2.1248 \\
& RHL&	ZnSnO$_3$                   & 245943 & 	0.254	&	D&1.2757 \\
& BCT&	SrSnO$_3$                   & 153532 & 	0.147	&	I&1.5157 \\
& RHL&	AgSbO$_3$                   & 245292 & 	0.240	&	I&1.1090  \\
& MCL&	Na$_2$Zn$_2$O$_3$   & 25617  &	0.160	&	D&1.8152 \\
& BCT&	CdIn$_2$O$_4$      & 52389  &	0.191	&	I&1.2032 \\
& BCT&	CaIn$_2$O$_4$      & 52390  &	0.312	&	D&2.1761 \\
& FCC&	MgIn$_2$O$_4$      & 157770  &	0.257	&	D&1.8340  \\
& FCC&	ZnGa$_2$O$_4$     & 81107  &	0.257	&	I&2.8600   \\
& FCC&	CdGa$_2$O$_4$     & 159739  &	0.202	&	I&1.5889 \\
& FCC&	MgGa$_2$O$_4$     & 86507  &	0.292	&	D&3.2937 \\
& FCC&	CdAl2$_2$O$_4$     & 183382  &	0.307	&	D&2.8039 \\
& FCC&	MgAl$_2$O$_4$       & 182859  &	0.376	&	D&5.1104 \\
& FCC&	ZnAl$_2$2O$_4$      & 94159  &	0.358	&	D&4.4947 \\
    \hline\hline
   \end{tabular}
   \end{center}
   \label{tab:binary_continued}
\end {table*}

\begin{figure}[htb]
\begin{center}
    \includegraphics[width=0.95\columnwidth]{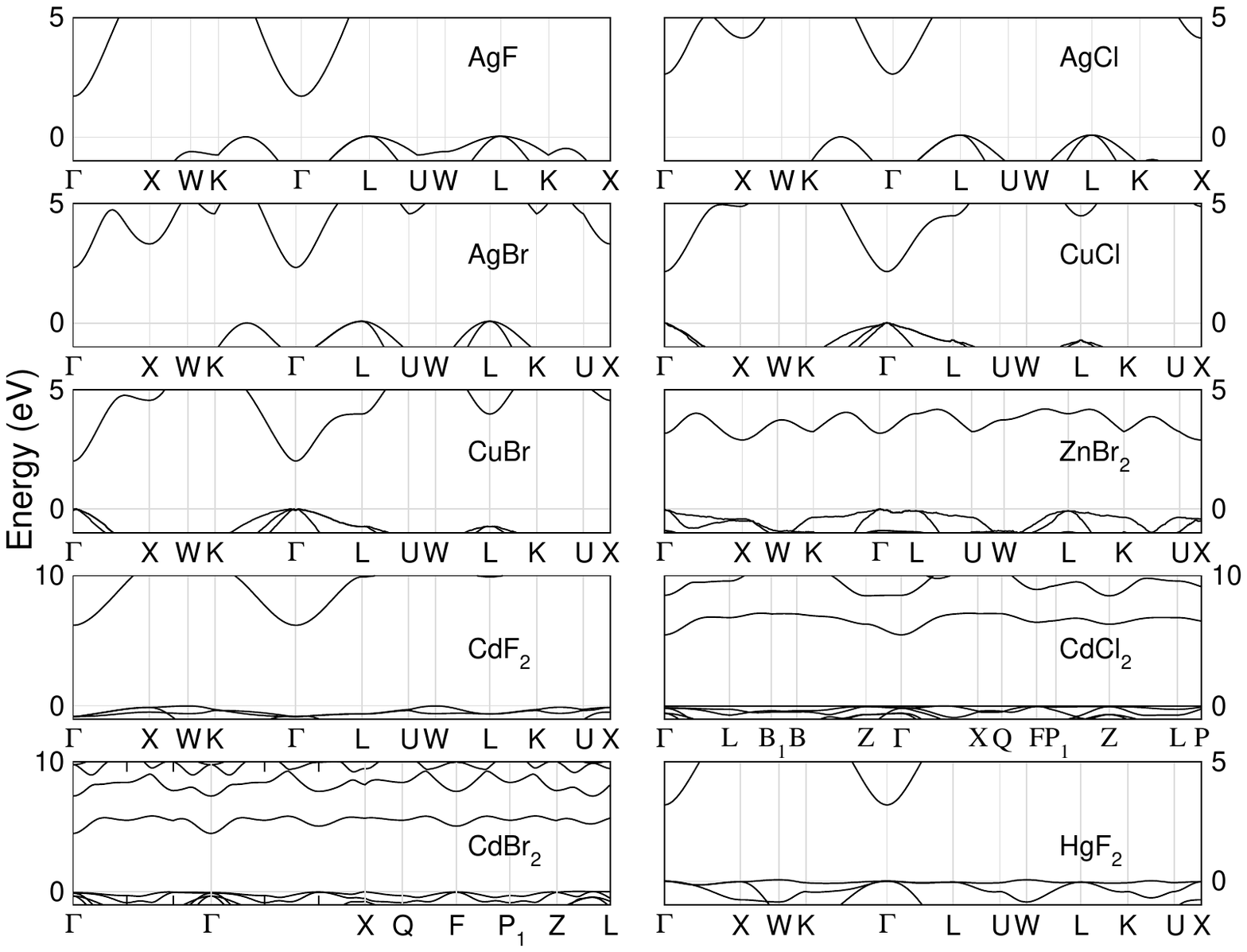}
    %\vspace{3mm}
    \caption{\small Band structures obtained with the Hubbard correction for the first set non-oxides hosts as listed in Table 1 of the main text. The zero of the energy is fixed at the maximum of the valence, the paths are the standard ones chosen from the {\em AFLOWLIB} repository and the energy window shown for each system is such that the TC characteristics are clearly visible.}
    \label{Bands_non_oxides_1_fig}
  \end{center}
\end{figure}

\begin{figure}[htb]
\begin{center}
    \includegraphics[width=0.95\columnwidth]{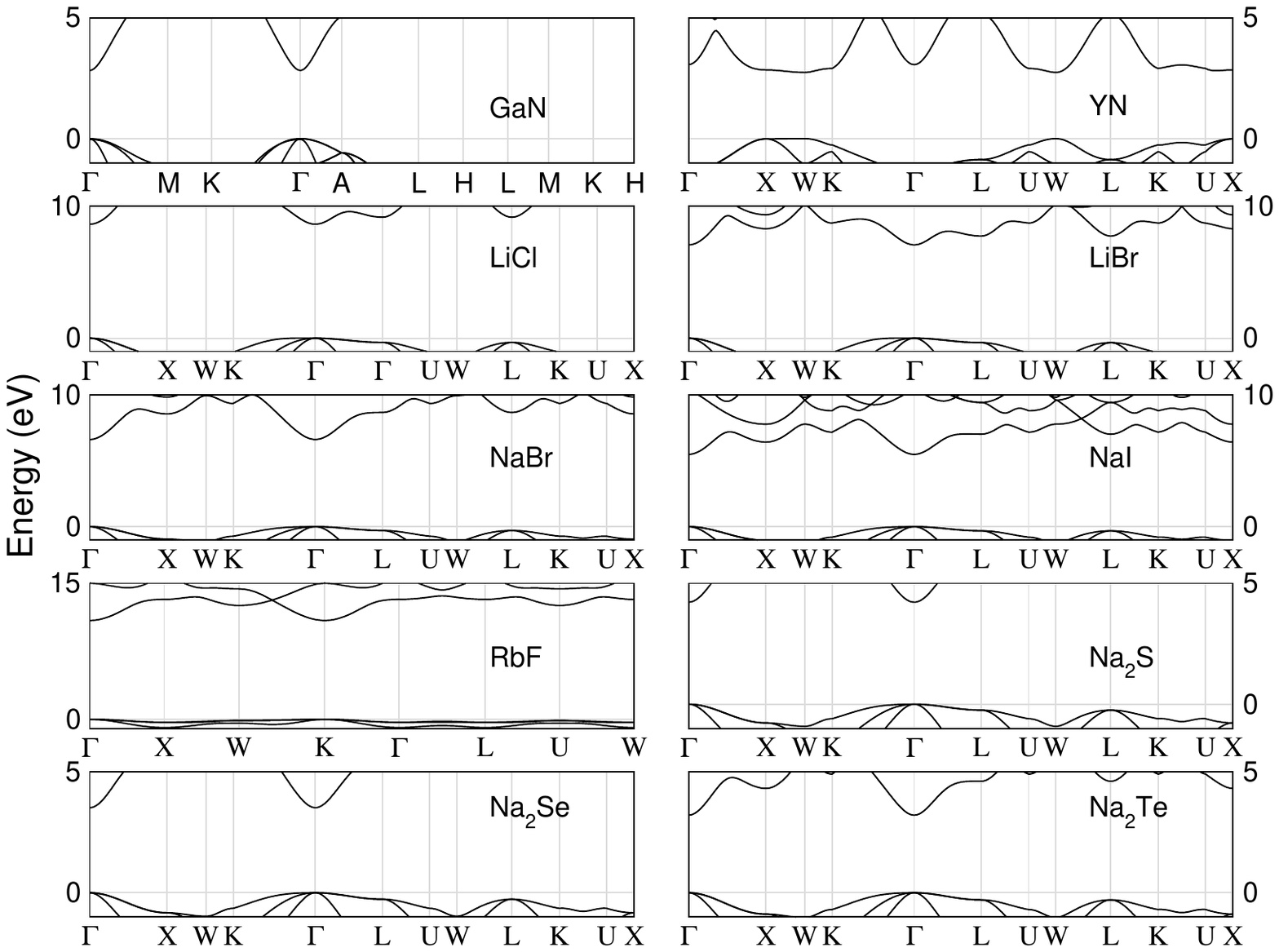}
    %\vspace{3mm}
    \caption{\small Band structures obtained with the Hubbard correction for the second set of non-oxides hosts as listed in Table 1 of the main text. The zero of the energy is fixed at the maximum of the valence, the paths are the standard ones chosen from the {\em AFLOWLIB} repository and the energy window shown for each system is such that the TC characteristics are clearly visible.}
    \label{Bands_non_oxides_2_fig}
  \end{center}
\end{figure}

\begin{figure}[htb]
\begin{center}
    \includegraphics[width=0.95\columnwidth]{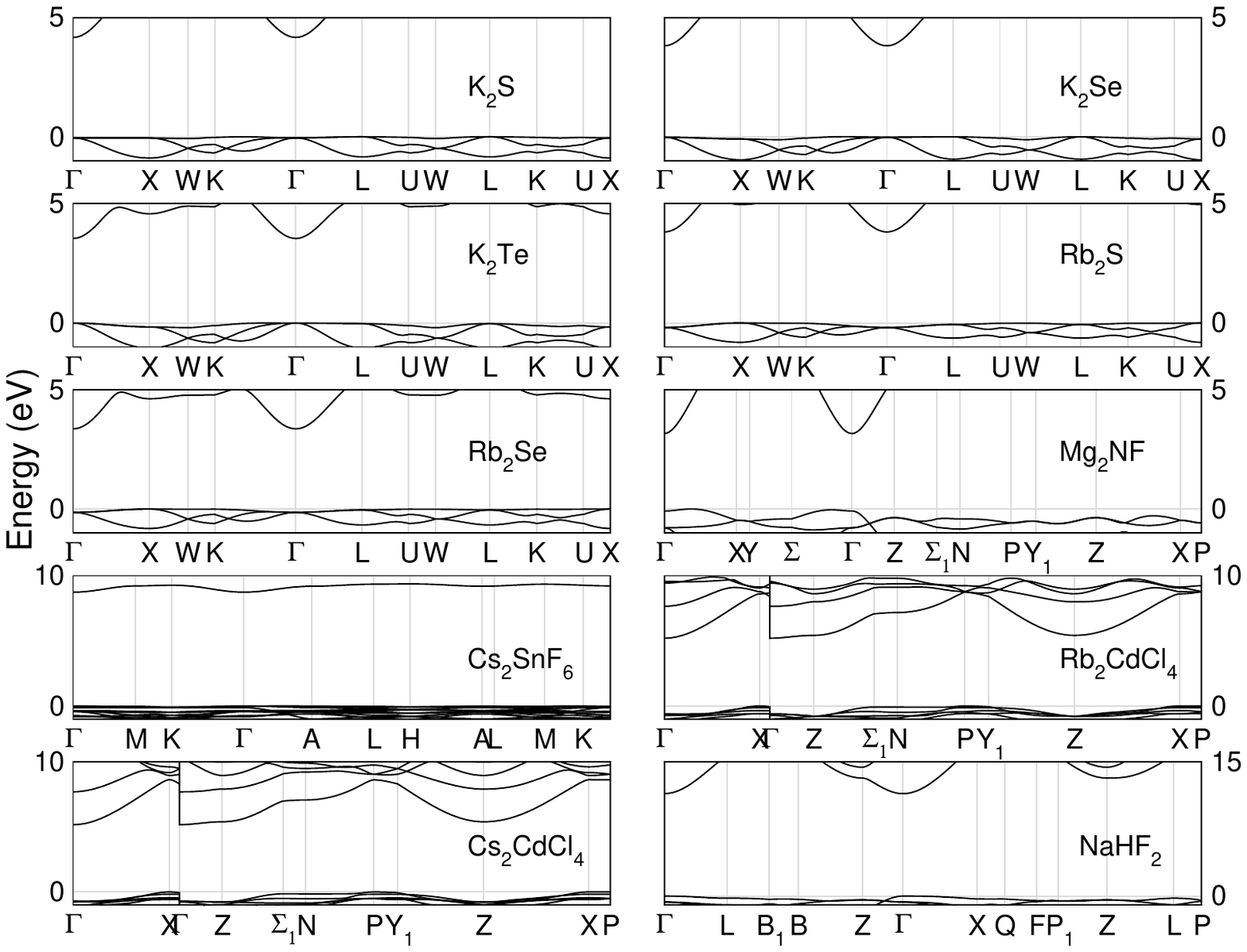}
    %\vspace{3mm}
    \caption{\small Band structures obtained with the Hubbard correction for the third set of non-oxides hosts as listed in Table 1 of the main text. The zero of the energy is fixed at the maximum of the valence, the paths are the standard ones chosen from the {\em AFLOWLIB} repository and the energy window shown for each system is such that the TC characteristics are clearly visible.}
    \label{Bands_non_oxides_3_fig}
  \end{center}
\end{figure}

\begin{figure}[htb]
\begin{center}
    \includegraphics[width=0.95\columnwidth]{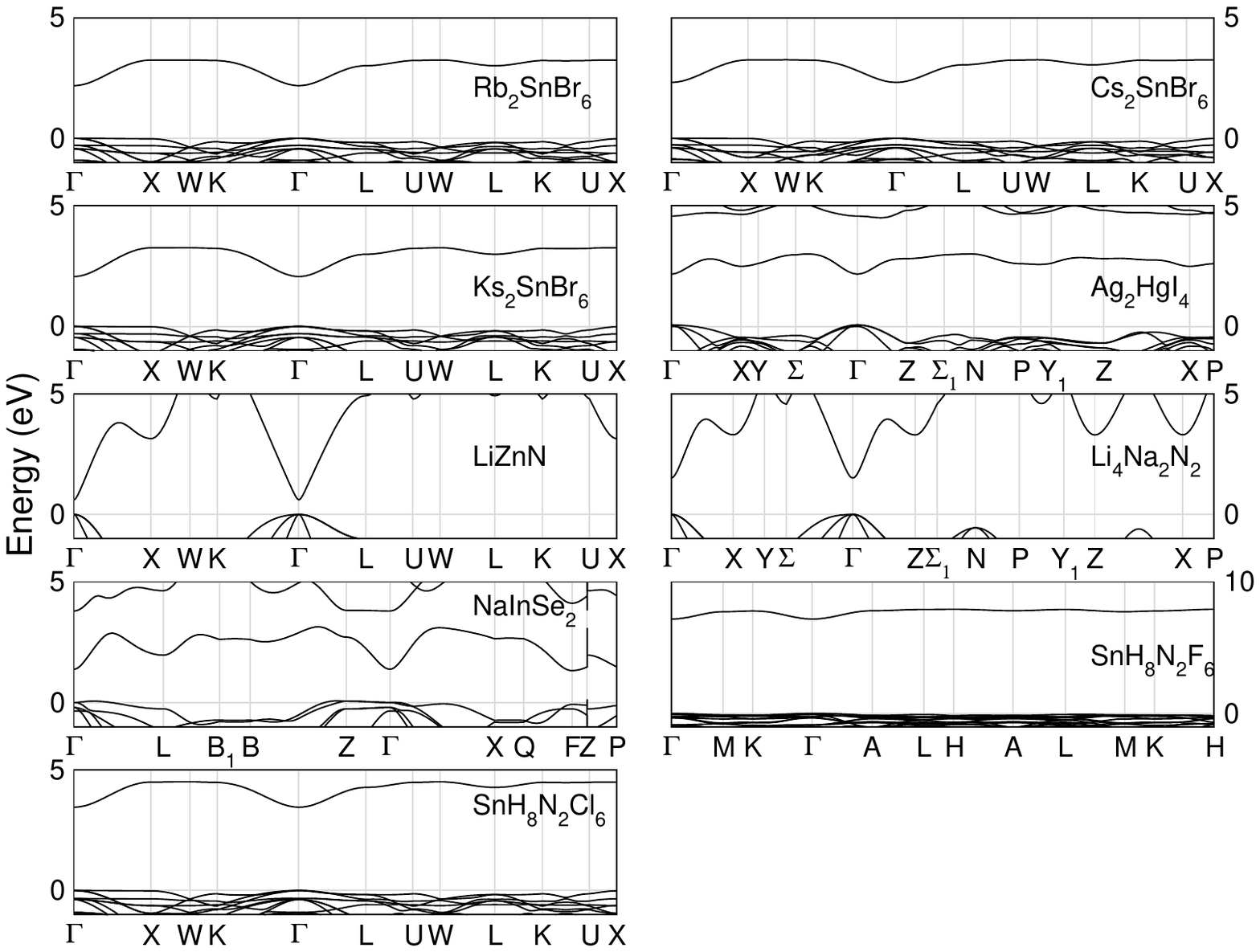}
    %\vspace{3mm}
    \caption{\small Band structures obtained with the Hubbard correction for the fourth set of non-oxides hosts as listed in Table 1 of the main text. The zero of the energy is fixed at the maximum of the valence, the paths are the standard ones chosen from the {\em AFLOWLIB} repository and the energy window shown for each system is such that the TC characteristics are clearly visible.}
    \label{Bands_non_oxides_4_fig}
  \end{center}
\end{figure}

\begin{table*}
\caption {Ternary compounds resulting from screenig the AFLOW database with the descriptors in Eqs.~(1-3) of the main text. The list contains information on  the symmetry group,  effective mass ($m^*/m_0$), gap type and its value $E_g^{Th}$ as founded in the {\em AFLOWLIB} repository (continued).}
\begin{center}
        {\bf{TERNARY COMPOUNDS}}\\
        \vspace{3mm}
    \begin{tabular}{ c|c|c|c|c|c|c  }
    \hline\hline
        & Structure & Material & ICSD & $m^*/m_0$ & Gap  & $E_g (eV)$ \\
    \hline\hline
$\bf{Oxides}$ & BCT&	Sr$_2$SnO$_4$        & 150388  &	0.388	&	I&2.1886 \\
& BCT&	Zn$_2$SiO$_4$        & 167188  &	0.347	&	I&3.3645 \\
& MCLC&	Cd$_2$H$_4$O$_4$ & 40186  &	0.265	&	I&1.8550  \\
& BCT&	Na$_6$PbO$_5$ & 15102  &	0.327	&	D&1.0991 \\
& HEX&	CdAs$_2$O$_6$ & 280576 & 	0.367	&	I&1.9731 \\
& HEX&	CdSb$_2$O$_6$ & 181929 & 	0.345	&	I&1.4613 \\
& MCLC&	In$_2$Ge$_2$O$_7$   & 74896  &	0.230	&	D&2.0144 \\
& MCLC&	Mg$_2$P$_2$O$_7$   & 22328  &	0.437	&	I&5.3739 \\
& MCLC&	Cd$_2$As$_2$O$_7$ & 280579 & 	0.266	&	I&1.7846 \\
& MCLC&	Ca$_2$As$_2$O$_7$ & 32602  &	0.470	&	I&3.3384 \\
& MCLC&	Mg$_2$As$_2$O$_7$ & 16885  &	0.403	&	D&3.0918 \\
& MCLC&	In$_2$Si$_2$O$_7$    & 74897  &	0.352	&	D&2.8794 \\
& HEX&	Rb$_2$In$_4$O$_7$  & 6321     &	0.123	&	I&1.6119 \\
& FCC&	Be$_4$TeO$_7$         & 1322     &	0.324	&	D&1.2844 \\
& RHL&	In$_6$TeO$_{12}$ & 245526 &	0.4284	&	I&1.1401 \\
\hline\hline
\bf{Non-oxides} & FCC&	Rb$_2$SnBr$_6$ 	& 158956 &	0.389	&	D&1.0987 \\
& FCC&	Cs$_2$SnBr$_6$ 	& 158957 &	0.461	&	D&1.2391 \\
& FCC&	K$_2$SnBr$_6$ 	& 158955  &	0.350	&	D&1.0046 \\
& HEX&	Cs$_2$SnF$_6$	 & 281  &	        0.193	&	D&5.3234 \\
& BCT&	Rb$_2$CdCl$_4$ 	& 51168  &	0.428	&	I&3.3283 \\
& BCT&	Cs$_2$CdCl$_4$ 	& 16576  &	0.442	&	I&3.3288 \\
& BCT&	Ag$_2$HgI$_4$ 	& 150343  &	0.289	&	D&1.5042 \\
& BCT&	Mg$_2$NF 		& 262327  &	0.306	&	I&2.1120  \\
& FCC&	LiZnN			 & 16790  	&	0.249	&	D&1.2871 \\
& CUB&	CsCdBr 			& 24483 	& 	0.357	&	I&1.1080  \\
& RHL&	NaHF$_2$ 		& 26870  &	0.439	&	I&6.8526 \\
& BCT&	Li$_4$Na$_2$N$_2$ & 92305  &	0.362	&	D&2.1739 \\	
& RHL&	NaInSe$_2$ 		& 25558 & 	0.064	&	I&1.0416 \\
\hline\hline	
   \end{tabular}
   \end{center}
\label{tab:ternary}
\end {table*}
\begin{table*}
\caption {Quaternary compounds resulting from screenig the AFLOW database with the descriptors in Eqs.~(1-3) of the main text. The list contains information on  the symmetry group,  effective mass ($m^*/m_0$), gap type and its value $E_g^{Th}$ as founded in the {\em AFLOWLIB} repository.}
\begin{center}
        {\bf{QUATERNARY COMPOUNDS}}\\
        \vspace{3mm}
    \begin{tabular}{ c|c|c|c|c|c|c  }
    \hline\hline
        & Structure & Material & ICSD & $m^*/m_0$ & Gap  & $E_g (eV)$ \\
    \hline\hline
\bf{Oxides} & BCT&	Sr$_2$GaSbO$_6$ 		& 157014  	& 0.332 	&	D&1.3090\\
& BCT&	KAg$_2$AsO$_4$ 		& 409793  	& 0.317	&	I&1.0831\\
& MCLC&	Na$_4$SrSi$_3$O$_9$		& 33943  		& 0.404	&	I&4.1273\\
& MCLC&	NaMgBO$_3$ 		& 249567  	& 0.491	&	D&3.6157\\
& MCLC&	Li$_3$Zn$_2$SbO$_6$		& 69189  		& 0.391	&	I&3.0659\\
& MCLC&	LiGaAs$_2$O$_7$ 		& 161500  	& 0.382	&	I&2.6197\\
& RHL&	SrSnB$_2$O$_6$ 		& 28267  		& 0.493	&	I&3.6321\\
& RHL&	NaCNO 		& 27138  		& 0.382	&	I&4.4500\\
& RHL&	KSbP$_2$O$_8$ 		& 61788  		& 0.375	&	I&2.9744\\
& FCC&	Ba$_2$YBiO$_6$ 		& 65555  		& 0.416	&	D&1.9822\\
& FCC&	Ba$_2$DyBiDyO$_6$		& 68612  		& 0.375	&	D&1.9038\\
& TET&	ClLuBi$_2$1O$_4$		& 92410 		& 0.452	&	D&1.3702	\\
\hline\hline
\bf{Non-oxides} &HEX&	SnH$_8$N$_2$F$_6$ 		& 409509  	& 0.343	&	D&5.1016\\
& FCC&	SnH$_8$N$_2$Cl$_6$		& 605  		& 0.488	&	D&2.0457\\
\hline\hline
   \end{tabular}
   \end{center}
\label{tab:quaternary}
\end {table*}
For the non-oxides hosts we report also in Figures (\ref{Bands_non_oxides_1_fig}, \ref{Bands_non_oxides_2_fig}, \ref{Bands_non_oxides_3_fig}, \ref{Bands_non_oxides_4_fig}) the band structures obtained with the inclusion of the Hubbard parameters U as discussed in the main text.
As can be seen from the figures, all the band structures show the wanted characteristics, namely: a large energy gap (that can be found listed in Table 1 of the main text), a clear separation in energy between the first and the second conduction band and a small effective mass associated to the curvature of the first conduction band. The bands are plotted along the standard paths as reported in the {\em AFLOWLIB} repository, the zero for each plot is fixed at the maximum of the last valence band and the energy window shown for each system is such that the TC characteristics are clearly visible.
\newpage

\newpage
%%%%%%%%%%%%%%%%%%%%%%%%%%%%%%%%
\section{Magnetic TC compounds}
%%%%%%%%%%%%%%%%%%%%%%%%%%%%%%%%

\begin{figure}[htb]
\begin{center}
    \includegraphics[width=0.95\columnwidth]{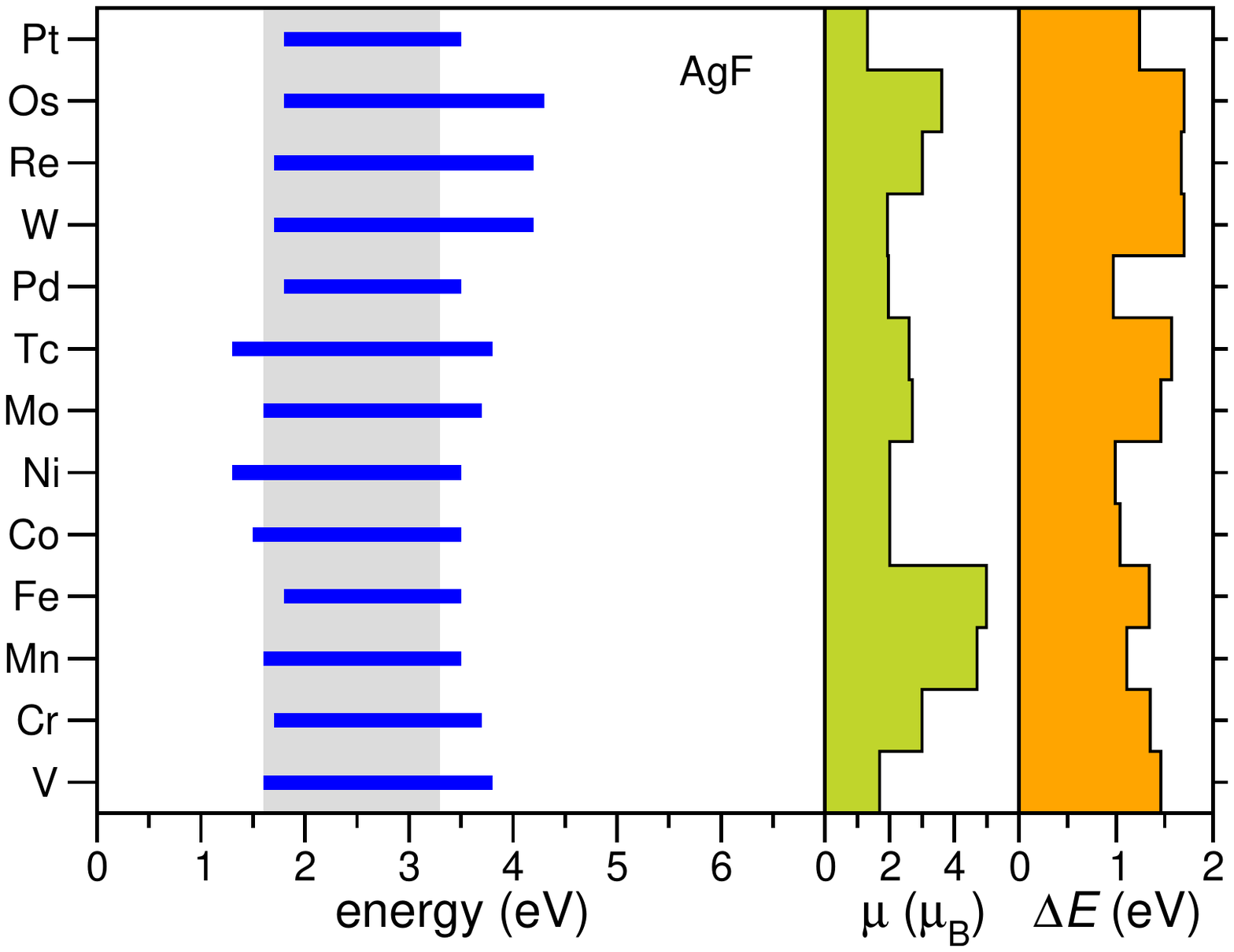}
    %\vspace{3mm}
    \caption{\small Energy distribution of the optical transmittance $T(E)$ (left panel), total magnetization $\mu$ (central panel),  and  $\Delta E$ (right panel) for TM:AgF and  (b) TM:CdCl$_2$ magnetic TCs. 
    Blue horizontal lines identify the transparency range $T(E)>0.9$; the shaded gray area indicates the visible range. Only the TM-host systems that satisfy the three conditions $\{\mathcal{C}_i\}$ described in the main text are reported.}
    \label{AgCl_fig}
  \end{center}
\end{figure}

\begin{figure}[htb]
\begin{center}
    \includegraphics[width=0.95\columnwidth]{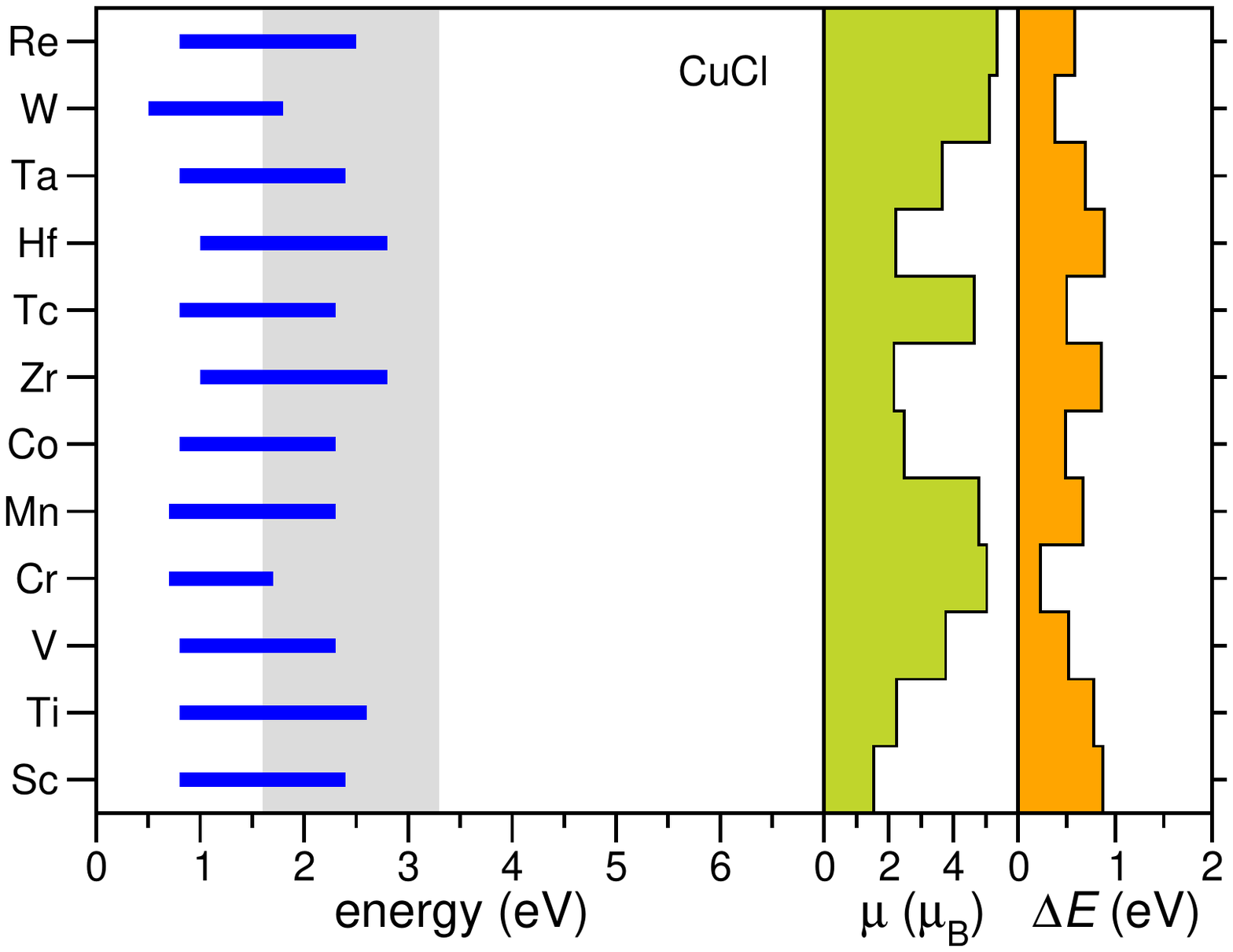}
    %\vspace{5mm}
    \caption{\small Energy distribution of the optical transmittance $T(E)$ (left panel), total magnetization $\mu$ (central panel),  and  $\Delta E$ (right panel) for TM:CuCl magnetic TCs. 
    Blue horizontal lines identify the transparency range $T(E)>0.9$; the shaded gray area indicates the visible range. Only the TM-host systems that satisfy the three conditions $\{\mathcal{C}_i\}$ described in the main text are reported.}
    \label{CuCl_fig}
  \end{center}
\end{figure}

\begin{figure}[htb]
\begin{center}
    \includegraphics[width=0.95\columnwidth]{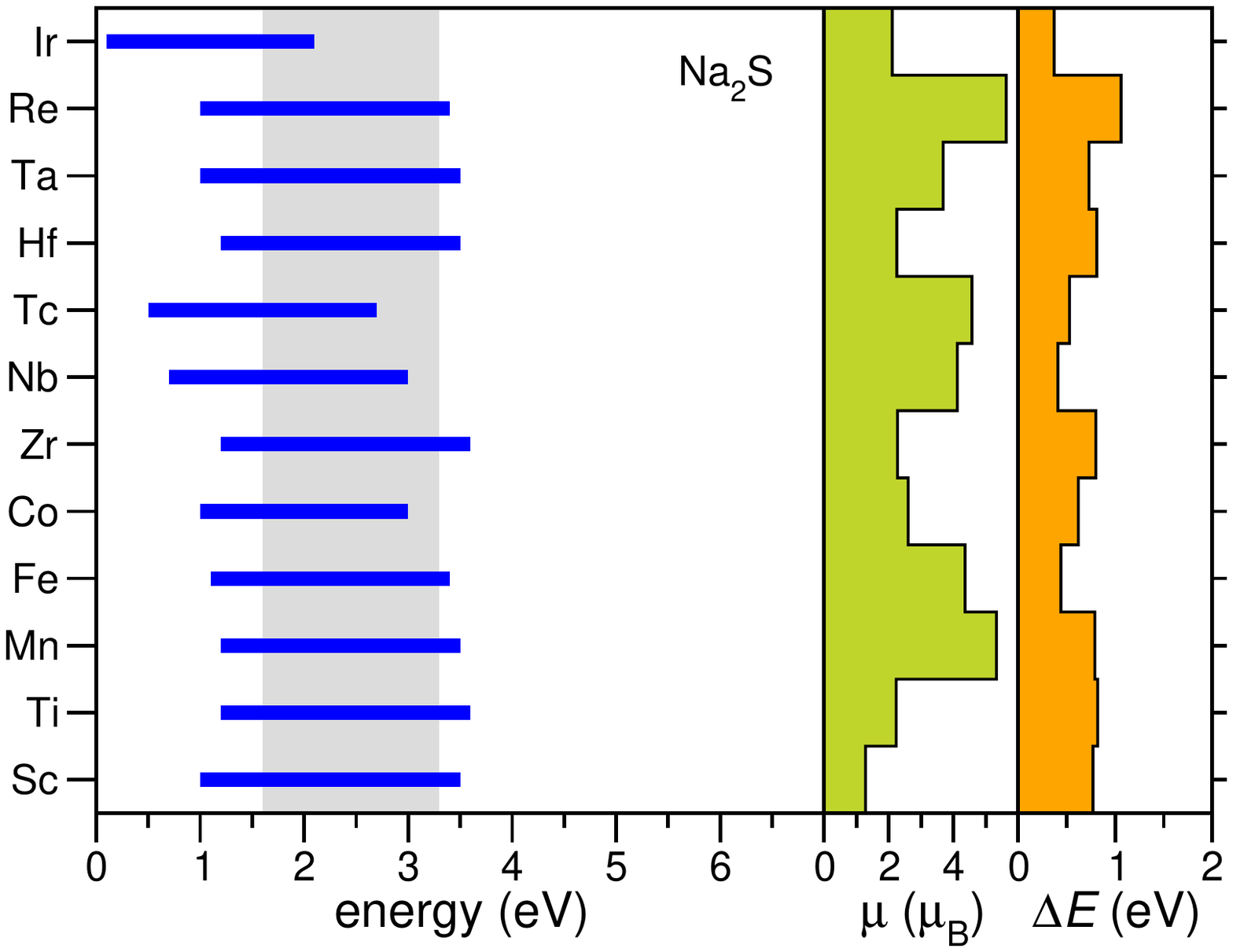}
    %\vspace{3mm}
    \caption{\small Energy distribution of the optical transmittance $T(E)$ (left panel), total magnetization $\mu$ (central panel),  and  $\Delta E$ (right panel) for TM:Na$_2$S magnetic TCs. 
    Blue horizontal lines identify the transparency range $T(E)>0.9$; the shaded gray area indicates the visible range. Only the TM-host systems that satisfy the three conditions $\{\mathcal{C}_i\}$ described in the main text are reported.}
    \label{Na2S_fig}
  \end{center}
\end{figure}

As shown in the main text, we have identified five possible representative non-oxides TC and for them we have performed an analysis exploring the effect of substitutional doping with TM elements. 
We have showed the result of such analysis in the main text for two candidates (AgF and CdCl$_2$) and we report here similar results for the three remaining magnetic-TC candidates (AgCl, CuCl and Na$_2$S).
In Figures (\ref{AgCl_fig}, \ref{CuCl_fig}, \ref{Na2S_fig}) we report the transparency window, the magnetization and the value of the Fermi energy measured with respect to the bottom of the conduction band ($\Delta$E) for the three compounds respectively.
We observe that AgCl and Na$_2$S have comparable behaviour with respect to AgF having a transparency window that covers almost enterely the visible spectrum and partially the IR (except for Ir-doped Na$_2$S and Cr-doped AgCl that are transparent almost only in the IR range).  
Differently CuCl shows a transparency window centered around the minimum of the visible spectrum and spanning a range between the IR and the visible spectrum.
